\documentclass[twocolumn]{aastex63}

\usepackage{amssymb,amsmath,amsthm}
\usepackage{graphicx}
\usepackage{mathrsfs}
\usepackage{booktabs}
\usepackage{multirow}
\usepackage{empheq}
\usepackage{verbatim}
\usepackage{mathtools}
\usepackage{hyperref}
\usepackage{xcolor}
\usepackage{xspace}

\newcommand{\uv}{$(u,v)$\xspace}        
\newcommand{\kms}{km\,s$^{-1}$\xspace}  
\newcommand{\kmsyr}{km\,s$^{-1}$\,yr$^{-1}$\xspace}  
\newcommand{\kmsmpc}{km\,s$^{-1}$\,Mpc$^{-1}$\xspace}   
\newcommand{\snr}{S/N\xspace} 

\shorttitle{Megamaser Cosmology Project XI}
\shortauthors{Pesce et al.}

\begin{document}

\title{The Megamaser Cosmology Project. XI. A geometric distance to CGCG 074-064}

\correspondingauthor{Dominic~W.~Pesce}
\email{dpesce@cfa.harvard.edu}

\author[0000-0002-5278-9221]{D.~W.~Pesce}
\affiliation{Center for Astrophysics $|$ Harvard \& Smithsonian, 60 Garden Street, Cambridge, MA 02138, USA}
\affiliation{Black Hole Initiative at Harvard University, 20 Garden Street, Cambridge, MA 02138, USA}

\author{J.~A.~Braatz}
\affiliation{National Radio Astronomy Observatory, 520 Edgemont Road, Charlottesville, VA 22903, USA}

\author{M. J. Reid}
\affiliation{Center for Astrophysics $|$ Harvard \& Smithsonian, 60 Garden Street, Cambridge, MA 02138, USA}

\author{J. J. Condon}
\affiliation{National Radio Astronomy Observatory, 520 Edgemont Road, Charlottesville, VA 22903, USA}

\author{F. Gao}
\affiliation{Key Laboratory for Research in Galaxies and Cosmology, Shanghai Astronomical Observatory, Chinese Academy of Science, Shanghai 200030, China}
\affiliation{National Radio Astronomy Observatory, 520 Edgemont Road, Charlottesville, VA 22903, USA}

\author{C. Henkel}
\affiliation{Max-Planck-Institut f\"ur Radioastronomie, Auf dem H\"ugel 69, D-53121 Bonn, Germany}
\affiliation{Astronomy Department, Faculty of Science, King Abdulaziz University, P.O. Box 80203, Jeddah 21589, Saudi Arabia}

\author{C. Y. Kuo}
\affiliation{Department of Physics, National Sun Yat-Sen University, No.70, Lianhai Rd., Gushan Dist., Kaohsiung City 804, Taiwan (R.O.C.)}

\author{K. Y. Lo}\altaffiliation{Deceased}
\affiliation{National Radio Astronomy Observatory, 520 Edgemont Road, Charlottesville, VA 22903, USA}

\author{W. Zhao}
\affiliation{Shanghai Observatory, 80 Nandan Road, Shanghai 200030, China}
\affiliation{Key Laboratory of Radio Astronomy, Chinese Academy of Sciences, 210008 Nanjing, PR China}

\begin{abstract}
As part of the survey component of the Megamaser Cosmology Project, we have discovered a disk megamaser system in the galaxy CGCG 074-064.  Using the GBT and the VLA, we have obtained spectral monitoring observations of this maser system at a monthly cadence over the course of two years.  We find that the systemic maser features display line-of-sight accelerations of $\sim$4.4\,\kmsyr that are nearly constant with velocity, while the high-velocity maser features show accelerations that are consistent with zero.  We have also used the HSA to make a high-sensitivity VLBI map of the maser system in CGCG 074-064, which reveals that the masers reside in a thin, edge-on disk with a diameter of $\sim$1.5\,mas (0.6\,pc).  Fitting a three-dimensional warped disk model to the data, we measure a black hole mass of $2.42^{+0.22}_{-0.20} \times 10^7$\,M$_{\odot}$ and a geometric distance to the system of $87.6^{+7.9}_{-7.2}$\,Mpc.  Assuming a CMB-frame recession velocity of $7308 \pm 150$\,\kms, we constrain the Hubble constant to $H_0 = 81.0^{+7.4}_{-6.9}$\,(stat.)\,$\pm 1.4$\,(sys.)\,\kmsmpc.
\end{abstract}

\section{Introduction}

Water megamasers residing in the accretion disks around supermassive black holes (SMBHs) provide a unique way to bypass the cosmic distance ladder and make one-step, geometric distance measurements to their host galaxies.  First applied to the archetypal megamaser-hosting galaxy NGC 4258 \citep{1999Natur.400..539H}, the ``megamaser technique'' exploits the simple geometry and Keplerian dynamics of maser clouds orbiting in a point-source potential to simultaneously solve for the SMBH mass and angular-size distance.  $H_0$ measurements made using the megamaser technique are independent of standard candles and the CMB, and thus provide a key piece of evidence for interpreting the current tension between early- and late-Universe measurements of $H_0$ \citep[see, e.g.,][]{Verde_2019}.

The Megamaser Cosmology Project (MCP) is a multi-year campaign to survey active galactic nuclei (AGN) for the presence of water megamasers, monitor their spectral evolution, and map their structure using very long baseline interferometry (VLBI).  The goal of the MCP is to determine $H_0$ with an accuracy of $\pm 3\%$ by making geometric distance measurements to megamaser galaxies in the Hubble flow \citep{2013ApJ...767..154R, 2013ApJ...767..155K, 2015ApJ...800...26K, 2016ApJ...817..128G}.

In this work we present a megamaser distance measurement to CGCG 074-064, a Seyfert 2 galaxy whose maser system was discovered in 2015 as part of the survey component of the MCP.  This paper is organized as follows.  In \autoref{ObservationsAndReduction} we describe the monitoring and mapping observations and data reduction procedures.  \autoref{Measurements} goes over our measurement techniques for determining maser positions and accelerations, and in \autoref{sec:HubbleConstant} we detail our modeling procedure and corresponding $H_0$ measurement.  \autoref{Discussion} discusses the observed VLBI continuum emission and spectral variability of the maser features.  Unless otherwise specified, all velocities referenced in this work use the optical definition in the barycentric reference frame.  The conversion from barycentric to CMB frame velocities is $v_{\text{CMB}} = v_{\text{bary}} + 263.3$\,\kms for CGCG 074-064 \citep{2009ApJS..180..225H}.

\section{Observations and data reduction} \label{ObservationsAndReduction}

There are two classes of observations necessary for making a Hubble constant measurement using a disk maser system: (1) high-sensitivity VLBI observations to map the spatial distribution of the masers, and (2) short-cadence ($\sim$monthly) monitoring observations spanning a sufficiently long time baseline to measure the accelerations of the systemic maser features.  We used the Very Long Baseline Array (VLBA), augmented with the Robert C. Byrd Green Bank Telescope (GBT) and the phased Karl G. Jansky Very Large Array (VLA), to map the maser system in CGCG 074-064.  The bulk of the monitoring spectra were taken with the GBT, though we used the VLA to observe during the summer months when the weather in Green Bank makes K-band observations inefficient.

\subsection{GBT monitoring observations} \label{GBTObs}

We performed $\sim$monthly spectral monitoring observations of CGCG 074-064 from 2015 October through 2017 May, for a total of 20 epochs (see \autoref{tab:MonitoringObservations}).  16 of the monitoring spectra were taken with the GBT.  Our general observing strategy and data reduction process follow similar procedures to those detailed in previous MCP papers \citep{2010ApJ...718..657B, 2015ApJ...810...65P}, so in this section we give only a brief overview.  All GBT data were reduced using GBTIDL\footnote{\url{http://gbtidl.nrao.edu/}}.

For each 3-hour GBT monitoring epoch we performed nodding observations with two of the seven beams of the K-band Focal Plane Array (KFPA), using the Versatile GBT Astronomical Spectrometer (VEGAS) as the backend.  The spectrometer was configured with four overlapping 187.5~MHz spectral windows, covering recession velocities from 3500--12500\,\kms contiguously with 5.722~kHz ($\sim$0.08\,\kms) spectral channels.  Both left circular polarization (LCP) and right circular polarization (RCP) were observed simultaneously in each of the two beams, and we performed hourly observations of a nearby bright ($>$1~Jy) continuum source to derive pointing and focus corrections.

During data reduction we smoothed the reference beam spectrum with a 64-channel boxcar function prior to differencing.  For every monitoring run, all integrations in both polarizations were averaged using a $\tau/T_{\text{sys}}^2$ weighting scheme (with $\tau$ the integration time and $T_{\text{sys}}$ the system temperature) chosen to minimize the final noise level.  A polynomial (typically third-order) was fit to line-free spectral channels and subtracted to remove any residual baseline structure from the final spectrum.  \autoref{tab:MonitoringObservations} lists the system temperatures and sensitivities achieved for all monitoring observations.

\begin{figure*}[t]
	\centering
		\includegraphics[width=1.00\textwidth]{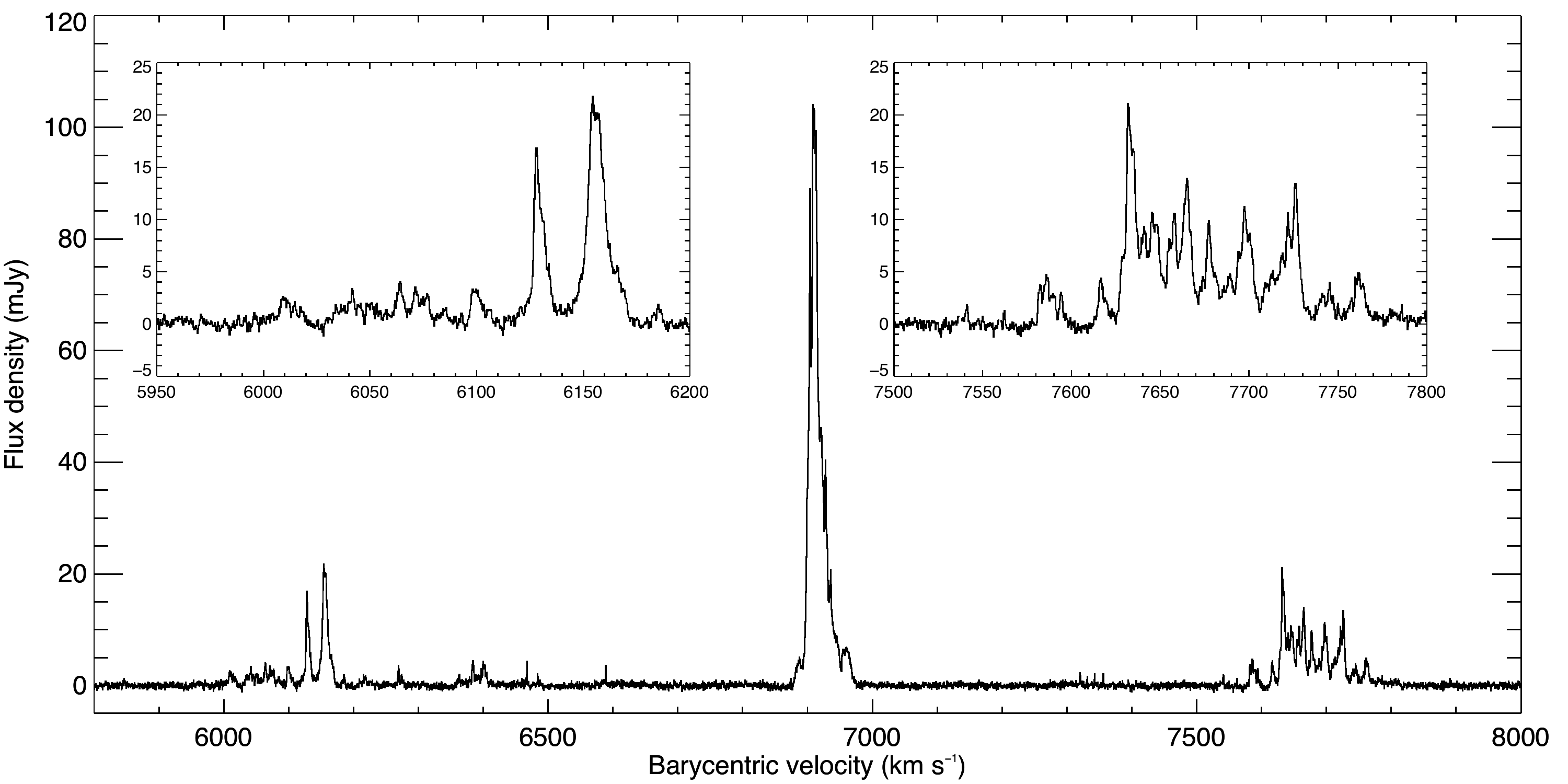}
	\caption{22 GHz GBT spectrum of CGCG 074-064, plotted as a weighted average over all epochs.  The two inset plots show zoomed-in spectra of the strongest high-velocity maser features.  The RMS noise level in this spectrum is 0.33~mJy per 0.3\,\kms spectral channel.}
	\label{fig:GBT_spectrum}
\end{figure*}

\autoref{fig:GBT_spectrum} shows the CGCG 074-064 maser spectrum averaged over all GBT epochs.  This spectrum represents the product of some $\sim$40 hours of integration, and it achieves an RMS noise level of 0.33~mJy per 0.32\,\kms spectral channel.  Maser emission is detected all the way down to the sensitivity limit, and individual maser features are seen out to velocity extremes of 7892\,\kms on the redshifted side and 5846\,\kms on the blueshifted side, corresponding to orbital velocities of $\sim$1000\,\kms.

\begin{deluxetable*}{lcccccc}
\tablecolumns{7}
\tablewidth{0pt}
\tablecaption{Monitoring observation details\label{tab:MonitoringObservations}}
\tablehead{	&		&		&	\colhead{$T_{\text{sys}}$}	&	\colhead{Sensitivity}	&	\colhead{Synthesized beam}	&	\colhead{Continuum flux density}	\\
\colhead{Epoch} & \colhead{Date} & \colhead{Telescope} &	\colhead{(K)}	&	\colhead{(mJy)}	&	\colhead{($\prime\prime \times \prime\prime$, $^{\circ}$)}	&	\colhead{($\mu$Jy)}}
\startdata
1		&	2015 Oct 15	&	GBT	&	42.1		&	3.8	&	\ldots											&	\ldots	\\
2		&	2015 Nov 13	&	GBT	&	44.4		&	2.9	&	\ldots											&	\ldots \\
3		&	2015 Dec 18	&	GBT	&	41.3		&	2.7	&	\ldots											&	\ldots \\
4		&	2016 Jan 12	&	GBT	&	45.9		&	2.9	&	\ldots											&	\ldots \\
5		&	2016 Feb 26	&	GBT	&	43.7		&	3.2	&	\ldots											&	\ldots \\
6		&	2016 Mar 22	&	GBT	&	47.8		&	2.7	&	\ldots											&	\ldots \\
7		&	2016 Apr 10	&	GBT	&	43.7		&	2.7	&	\ldots											&	\ldots \\
8		&	2016 Jun 08	&	VLA	&	\ldots	&	1.3	&	$0.38 \times 0.31$, $-2.79$	&	$62.1 \pm 6.0$ \\
9		&	2016 Jun 13	&	GBT	&	49.5		&	3.2	&	\ldots											&	\ldots	\\
10	&	2016 Jul 10	&	VLA	&	\ldots	&	2.8	&	$0.38 \times 0.34$, $17.12$	&	$83.3 \pm 12.8$	\\
11	&	2016 Aug 15	&	VLA	&	\ldots	&	3.0	&	$0.36 \times 0.33$, $45.05$	&	$90.7 \pm 11.7$	\\
12	&	2016 Sep 10	&	VLA	&	\ldots	&	2.1	&	$0.25 \times 0.15$, $80.69$	&	$58.9 \pm 8.7$	\\
13	&	2016 Oct 09	&	GBT	&	47.6		&	3.0	&	\ldots											&	\ldots	\\
14	&	2016 Nov 18	&	GBT	&	50.4		&	3.1	&	\ldots											&	\ldots	\\
15	&	2016 Dec 14	&	GBT	&	35.2		&	2.2	&	\ldots											&	\ldots	\\
16	&	2017 Jan 25	&	GBT	&	50.3		&	3.2	&	\ldots											&	\ldots	\\
17	&	2017 Feb 16	&	GBT	&	37.7		&	2.5	&	\ldots											&	\ldots	\\
18	&	2017 Mar 15	&	GBT	&	40.0		&	2.7	&	\ldots											&	\ldots	\\
19	&	2017 Apr 11	&	GBT	&	64.7		&	4.3	&	\ldots											&	\ldots	\\
20	&	2017 May 09	&	GBT	&	42.3		&	2.8	&	\ldots											&	\ldots	\\
\enddata
\tablecomments{Monitoring observation details.  The sensitivity is listed per 5.722~kHz (0.08\,\kms) channel for all GBT observations and per 15.625~kHz (0.21\,\kms) channel for all VLA observations.  The RMS sensitivity for each epoch was determined using the line-free velocity range spanning 7100--7300\,\kms.  The synthesized beam sizes are quoted as the full width at half maximum (FWHM) of the major $\times$ minor axes of the restoring elliptical Gaussian in the 6900\,\kms channel, with position angles measured east of north.  The continuum flux densities measured for the VLA tracks are quoted as the peak value of the unresolved point source measured in a $\sim$2~GHz bandwidth centered at a rest-frame frequency of 22.2~GHz.}
\end{deluxetable*}

\subsection{VLA monitoring observations} \label{VLAObs}

We used the VLA to observe during the 2016 summer months, when the level of atmospheric water vapor at the Green Bank site would have made K-band observations inefficient.  In total, four 3-hour tracks were covered by the VLA (see \autoref{tab:MonitoringObservations}).  All VLA data were reduced and imaged using standard procedures within CASA\footnote{\url{https://casa.nrao.edu/}}.

The first three tracks were observed with the VLA in B configuration, and the September track was observed while the VLA was transitioning between B and A configurations\footnote{We note that the August and September VLA tracks were observed during the period of time in semester 16B when the online tropospheric delay model was misapplied.  Though our observations are negligibly affected by this issue, we nevertheless applied a tropospheric delay error correction during the data reduction procedure for these tracks.}.  We configured the correlator to place three overlapping 64~MHz windows covering the three sets of maser features, with 4096 15.625~kHz (0.21\,\kms) channels in each spectral window.  An additional eight 128~MHz spectral windows with coarser (2~MHz; 26.5\,\kms) channel resolution were placed on each side of the maser profile, resulting in a net $\sim$2~GHz increase in bandwidth for a significant improvement in continuum sensitivity.

During reduction, we first corrected for antenna positions and atmospheric opacity before solving for delay and phase solutions on the flux calibrator (3C~286, which also doubled as our bandpass calibrator).  The brightest systemic maser features exceeded $\sim$150~mJy for all VLA tracks, and by averaging over a 10\,\kms window in both polarizations we were able to track the phase solutions on individual baselines with a two-minute cadence.  After applying the bandpass, flux, and phase calibrations we performed a round of (typically minor) data flagging and repeated the calibration procedure once more before splitting out the calibrated science target.  We then performed a series of phase and amplitude self-calibration steps, once again using the brightest systemic maser features.  We stopped iterating self-calibration once there was no noticeable increase in the signal-to-noise ratio (\snr), which typically occurred after 2-3 rounds.

Prior to imaging, we performed continuum subtraction on the \uv-data.  We imaged the continuum and spectral line cubes separately, using the \texttt{CLEAN} algorithm with natural \uv-weighting for both.  The continuum is unresolved in our VLA observations, and it is spatially coincident with the maser emission.  Combining all VLA tracks, we measure an average continuum level of $72.3 \pm 4.8$~$\mu$Jy across the 2~GHz bandwidth (centered at a rest-frame frequency of 22.2~GHz).  The continuum level shows strong (greater than $\sim$50\%) variability from one epoch to another, with a similar magnitude and timescale to that seen in the nuclear continuum emission from the megamaser galaxy NGC 4258 \citep{1997ApJ...475L..17H}.  Unlike in NGC 4258, we do not find evidence for a correlation between the continuum level and the average flux density of any group of maser features.

\subsection{VLBI mapping observations} \label{VLBIObs}

In total, we observed 10 6-hour VLBI tracks (see \autoref{tab:VLBI_Observations}).  The first track was phase-referenced to measure the absolute position of CGCG 074-064, while the subsequent 9 tracks were self-calibrated on the strongest systemic maser features (see \autoref{tab:Positions}).  As with the single-dish monitoring observations we have generally followed the same observing and data reduction procedures used for previous MCP targets \citep{2009ApJ...695..287R, 2011ApJ...727...20K, 2016ApJ...817..128G}, so this section focuses primarily on differences from previous MCP papers.  All VLBI data were reduced in AIPS\footnote{\url{http://www.aips.nrao.edu/}} and imaged with CASA.

For the phase-referenced track, we observed using only the VLBA antennas.  The correlator was configured with two overlapping 128~MHz spectral windows placed to either side of the systemic features.  Both windows contained the systemic complex of maser features, with one window shifted blueward and the other shifted redward to cover the high-velocity maser features.  Each spectral window was spanned by 256 channels spaced contiguously every 0.5~MHz ($\sim$6.7\,\kms), and we observed in dual circular polarization.  We used J1410+0731 as our phase-reference calibrator (separated from CGCG 074-064 by 2.3 degrees), switching between target and calibration observations on a 3-minute duty cycle.  We observed J1415+1320 hourly as a delay calibrator, and the entire track was bracketed by ``geodetic'' observations (see \citealt{2009ApJ...695..287R}).  We measure the absolute position of the maser system, defined as the intensity-weighted mean position of all systemic maser features, to be $\alpha_{\text{J2000}} = 14$:$03$:$04.457746$, $\delta_{\text{J2000}} = +08$:$56$:$51.03483$.   The resulting statistical and relative calibration uncertainties are much smaller than the absolute astrometric uncertainties for the phase-reference source, so we take the absolute positional uncertainties for CGCG 074-064 to be 0.78~mas in right ascension and 1.15~mas in declination (\autoref{tab:Positions}).

The self-calibrated tracks were observed using the High Sensitivity Array (HSA), composed of the VLBA plus the GBT and phased-VLA.  We used the same correlator configuration as for the phase-referenced track, but in addition we obtained a second ``zoom'' correlator pass with a higher-resolution channel spacing of 25~kHz ($\sim$0.34\,\kms) across three 64~MHz spectral windows contiguously covering the three sets of maser features.  As with the phase-referenced track, we performed hourly delay calibration observations of J1415+1320.  The VLA was ``phased-up'' every 10 minutes by observing J1351+0830, located 2.9 degrees away from CGCG 074-064.  Each track was bracketed by observations of either 4C39.25 or 3C286, which served as both fringe-finders and bandpass calibrators.  During data processing the strongest systemic features, located between 6900\,\kms and 6920\,\kms, were used to self-calibrate the phases.

After calibration we concatenated all of the phase-referenced tracks into one measurement set, weighting each track by its RMS (i.e., using $1/\sigma^2$ weighting).  We then imaged the dataset in CASA, using the \texttt{CLEAN} algorithm with natural \uv-weighting.  The RMS of the final data cube is 0.49~mJy~beam$^{-1}$ in a single $\sim$0.34\,\kms channel.  Prior to mapping the maser system, we averaged to $\sim$2\,\kms channels, corresponding to a typical maser feature linewidth \citep[e.g.,][]{2013ApJ...767..154R}.

\begin{figure}[t]
	\centering
		\includegraphics[width=1.00\columnwidth]{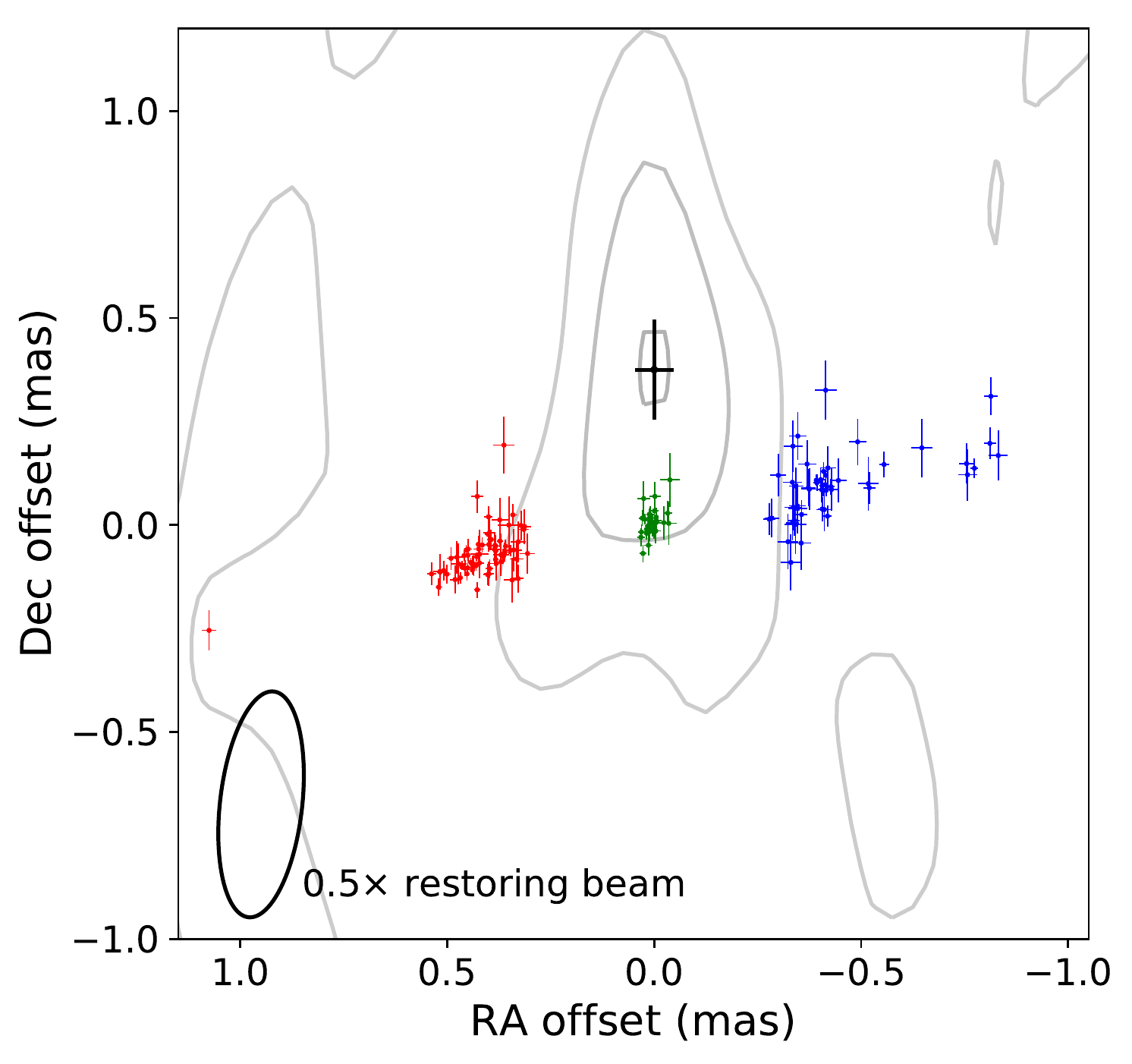}
	\caption{VLBI map of the maser system in CGCG 074-064, with maser spot positions extracted from the data cube as described in \autoref{PositionFitting}.  Maser spots are plotted with 1$\sigma$ uncertainties in right ascension and declination shown as horizontal and vertical lines, respectively.  Blue, green, and red points mark blueshifted, systemic, and redshifted masers, respectively.  The gray contours show the 22~GHz continuum, imaged using Briggs weighting with the robust parameter set to zero; the location of the continuum peak is marked in black with its associated $1\sigma$ uncertainties.  The continuum emission peaks at a value of ${\sim}46$~$\mu$Jy~beam$^{-1}$, and contours are shown at 15, 30, and 45~$\mu$Jy~beam$^{-1}$.  Note that the apparent North-South elongation of the continuum structure is driven by the asymmetric resolution element; the FWHM restoring beam, scaled down by a linear factor of 2, is shown at the bottom left-hand corner.}
	\label{fig:map_and_continuum}
\end{figure}

We imaged the line-free channels in our combined VLBI data using Briggs \uv-weighting with the robust parameter set to zero \citep{Briggs_1995}, and we detected a marginally-resolved continuum source with a peak flux density of $46 \pm 9.5$~$\mu$Jy~beam$^{-1}$ (see \autoref{fig:map_and_continuum}).  The peak of the continuum emission is located $0.38 \pm 0.12$~mas north of the disk plane, and it is aligned in right ascension with the systemic features.  The VLBI continuum is somewhat weaker (up to a factor of $\sim$2) than what was observed with the VLA, which could be explained by either the presence of an intermediate-scale component that is resolved out on very long baselines, or by source variability, which we know from the VLA monitoring observations is large enough to plausibly account for the entirety of the flux difference.

\begin{deluxetable*}{lcccc}
\tablecolumns{5}
\tablewidth{0pt}
\tablecaption{VLBI observation details\label{tab:VLBI_Observations}}
\tablehead{	&		&		&	\colhead{Synthesized beam}	&	\colhead{Sensitivity}  \\
\colhead{Project code} & \colhead{Date} & \colhead{Stations} &	\colhead{($\text{mas} \times \text{mas}$, $^{\circ}$)}	&	\colhead{(mJy)}}
\startdata
BB370Z$^a$	&	2016 Jan 19			&	VLBA								&	$2.33 \times 0.36$, $-18.18$								&	\hphantom{$^b$}1.30$^b$	\\
BB370D	&	2016 Feb 11			&	VLBA$+$GBT$+$VLA		&	$1.29 \times 0.47$, $-3.5$\hphantom{$00$}		&	\hphantom{$^c$}1.44$^c$	\\
BB370E	&	2016 Feb 21			&	VLBA$+$GBT$+$VLA		&	$1.05 \times 0.36$, $-6.88$\hphantom{$0$}		&	1.26	\\
BB370G	&	2016 Feb 28			&	VLBA$+$GBT$+$VLA		&	$1.48 \times 0.34$, $-14.3$\hphantom{$0$}		&	1.01	\\
BB370H	&	2016 Mar 10			&	VLBA$+$VLA$^d$			&	$1.40 \times 0.35$, $-17.24$								&	2.09	\\
BB370J	&	2016 Mar 21			&	VLBA$+$GBT$+$VLA		&	$1.28 \times 0.36$, $-9.63$\hphantom{$0$}		&	0.95	\\
BB370L	&	2016 Mar 24			&	VLBA$+$GBT$+$VLA		&	$1.29 \times 0.36$, $-14.26$								&	1.07	\\
BB370U	&	2016 May 16			&	VLBA$+$GBT$+$VLA		&	$1.13 \times 0.36$, $-7.4$\hphantom{$00$}		&	1.01	\\
BB370Y	&	2016 Jun 17/18	&	VLBA$+$GBT$+$VLA		&	$1.02 \times 0.36$, $-6.30$\hphantom{$0$}		&	\hphantom{$^e$}1.29$^e$	\\
BB370AB	&	2016 Jun 19/20	&	VLBA$+$GBT$+$VLA		&	$1.31 \times 0.34$, $-11.8$\hphantom{$0$}		&	1.76	 \\
\midrule
$\ldots$	&	$\ldots$				&	$\ldots$&	$1.12 \times 0.40$, $-6.91$\hphantom{$0$}		&	0.49 \\
\enddata
\tablecomments{VLBI observation details.  All tracks were 6 hours in length.  The RMS sensitivity for each track was determined using the line-free velocity range spanning 7100--7300\,\kms.  All tracks prior to BB370U were taken with the VLA in C-configuration, while all subsequent tracks had the VLA in B-configuration.  The synthesized beam sizes are quoted as the FWHM of the major $\times$ minor axes of the restoring elliptical Gaussian, with position angles measured east of north. The bottom row gives the beam characteristics and sensitivity from combining all tracks. \\
$^a$Track BB370Z was observed in a phase-referencing observing mode; all other tracks used self-calibration. \\
$^b$The sensitivity in the phase-referenced track is calculated per 0.5 MHz ($\sim$6.7\,\kms) channel from the default ``continuum-like'' correlator pass (i.e., without re-correlating at finer spectral resolution). \\
$^c$The sensitivity in the self-calibrated tracks is calculated per 25 kHz ($\sim$0.34\,\kms) channel from a second ``zoom'' correlator pass. \\
$^d$No fringes were found on GBT baselines for this track. \\
$^e$The GBT had poor pointing corrections for the first $\sim$2 hours of this track, so all GBT baselines were flagged during this time period.}
\end{deluxetable*}

\begin{deluxetable*}{lccccc}
\tablecolumns{6}
\tablewidth{0pt}
\tablecaption{VLBI positions for CGCG 074-064 and calibrators\label{tab:Positions}}
\tablehead{	&	\colhead{R.A.}	&	\colhead{decl.}	&	\colhead{Uncertainty in R.A.}	&	\colhead{Uncertainty in decl.}	&	 \\
\colhead{Name} &	\colhead{(J2000)}	&	\colhead{(J2000)}	&	\colhead{(mas)}	&	\colhead{(mas)}	& \colhead{Purpose}}
\startdata
4C39.25				&	09:27:03.013938		&	$+$39:02:20.85177	&	0.13	&	0.10	&	fringe finder/bandpass calibrator \\
3C286					&	13:31:08.288051		&	$+$30:30:32.95925	&	0.17	&	0.17	&	fringe finder/bandpass calibrator \\
J1351+0830		&	13:51:16.919081		&	$+$08:30:39.90354	&	0.09	&	0.19	&	VLA ``phase-up'' calibrator \\
J1410+0731		&	14:10:35.075347		&	$+$07:31:21.48972	&	0.78	&	1.15	&	phase reference calibrator \\
J1415+1320		&	14:15:58.817511		&	$+$13:20:23.71291	&	0.02	&	0.04	&	delay calibrator \\
\midrule
CGCG 074-064	&	14:03:04.457746		&	$+$08:56:51.03483	&	0.78	&	1.15	&	science target \\
\enddata
\tablecomments{VLBI positions for CGCG 074-064 and calibrators.  The positions for the calibrators are from the VLBA Calibrator Survey, and the position for CGCG 074-064 is measured in reference to J1410+0731.  The astrometric uncertainty in the phase reference calibrator J1410+0731 dominates the absolute position uncertainty for CGCG 074-064.}
\end{deluxetable*}

\section{Measurements} \label{Measurements}

The input data for our disk modeling consists of an on-sky position $(x,y)$, a line-of-sight velocity $v$, and a line-of-sight acceleration $a$ for each maser ``spot'' (i.e., for each velocity channel in the VLBI map).  In this section we detail how the maser positions and accelerations are measured.  \autoref{tab:Measurements} lists all measured (and some modeled) quantities for each maser spot.

\begin{deluxetable*}{lcccccccccc}
\tablecolumns{11}
\tablewidth{0pt}
\tablecaption{Measurements for individual maser spots\label{tab:Measurements}}
\tablehead{\colhead{Spot}	&	\colhead{Velocity}	&	\colhead{$S_{\nu}$}	&	\colhead{$\sigma_S$}	&	\colhead{$x$}	&	\colhead{$\sigma_x$}	&	\colhead{$y$}	&	\colhead{$\sigma_y$}	&	\colhead{$a$}	&	\colhead{$\sigma_a$}	&	\colhead{Accel.} \\
\colhead{type}	&	\colhead{(\kms)}	&	\colhead{(Jy)}	&	\colhead{(Jy)}	&	\colhead{(mas)}	&	\colhead{(mas)}	&	\colhead{(mas)}	&	\colhead{(mas)}	&	\colhead{(\kmsyr)}	&	\colhead{(\kmsyr)}	&	\colhead{meas.}}
\startdata
b	&	6007.40	&	0.00338	&	0.00022	&	$-$0.277944							&	0.012987	&	\hphantom{$-$}0.014065	&	0.036364	&	\hphantom{$-$}0.082	&	1.758	&	0	\\
b	&	6009.41	&	0.00252	&	0.00020	&	$-$0.284287							&	0.015984	&	\hphantom{$-$}0.016585	&	0.044755	&	\hphantom{$-$}0.020	&	2.164	&	0	\\
s	&	6897.66	&	0.00435	&	0.00017	&	\hphantom{$-$}0.011185	&	0.007604	&	$-$0.015560							&	0.021290	&	\hphantom{$-$}4.580	&	1.030	&	1	\\
s	&	6899.68	&	0.00519	&	0.00021	&	$-$0.000542							&	0.007909	&	\hphantom{$-$}0.033921	&	0.022144	&	\hphantom{$-$}4.140	&	1.071	&	1	\\
r	&	7650.97	&	0.00867	&	0.00024	&	\hphantom{$-$}0.452579								&	0.005421	&	$-$0.118197							&	0.015180	&	$-$0.106						&	0.734	&	1	\\
r	&	7652.97	&	0.00722	&	0.00018	&	\hphantom{$-$}0.450337								&	0.004854	&	$-$0.073305							&	0.013590	&	\hphantom{$-$}0.285	&	0.657	&	1	\\
\enddata
\tablecomments{Measurements for individual maser spots.  The ``spot type'' column 1 indicates which velocity group the maser spot belongs to (``b'' for blueshifted, ``s'' for systemic, ``r'' for redshifted).  The velocities in column 2 are quoted using the optical convention in the barycentric reference frame.  Columns 3 and 4 list the maser flux density and RMS from the VLBI channel maps.  Columns 5 through 8 list the position measurements and associated uncertainties.  Column 9 lists either the measured or modeled acceleration for each maser spot, and column 10 lists the associated uncertainties obtained from the disk modeling.  Column 11 indicates whether the acceleration for the maser spot was measured (``1'') or modeled (``0''). \\
(Only a portion of the table is shown here to illustrate its form and content.  This table is available in its entirety in machine-readable form.)}
\end{deluxetable*}

\subsection{Position fitting for the maser spots} \label{PositionFitting}

Even at the ${\lesssim}1$\,mas angular resolution afforded by VLBI, individual masers are unresolved point sources.  In any single velocity channel of a \texttt{CLEAN}ed image, a maser ``spot'' thus takes on the appearance of the restoring beam.  This beam is a two-dimensional (2D) elliptical Gaussian of known dimensions and position angle determined from the \uv-coverage of the observation and the weighting scheme used during the \texttt{CLEAN}ing process (see \autoref{tab:VLBI_Observations}), so every maser spot in the data cube will necessarily share these characteristics\footnote{There is a small (at the level of ${\sim}10^{-3}$ for CGCG 074-064) frequency-dependent gradient in the beam size across the maser spectrum, but this effect is also determined by the \uv-coverage and doesn't modify the beam shape.}.  The only unknown parameters for any given maser spot are then the centroid (i.e., the coordinate location in right ascension and declination of the Gaussian) and the amplitude.

We used a least-squares fitting routine \citep{2009ASPC..411..251M} to determine the amplitude and centroid of any maser spot within each velocity channel.  The fitted model was a 2D elliptical Gaussian with major axis, minor axis, and position angle fixed to match the restoring beam parameters.  Initial guesses for the centroid and amplitude were obtained using the location and value of the brightest pixel in each channel, and converged fits had typical reduced-$\chi^2$ values of $\sim$1.  Our resulting VLBI map is shown in \autoref{fig:map_and_continuum}.

For an image containing only a 2D elliptical Gaussian and some normally-distributed noise, we follow \cite{Condon_1997} and define the measured \snr to be the amplitude of the best-fit Gaussian divided by the RMS within a beam -- i.e., the standard deviation of the pixel values far from the peak of the Gaussian multiplied by the effective number of pixels contained within the beam area -- as measured from the signal-free regions of the image.  If we fit such an image using the model described above, the uncertainty in a measurement of one of the centroid coordinates ($\sigma_x$) will be related to the full width at half maximum (FWHM, $\Delta_x$) of the restoring beam along that direction by (see, e.g., \citealt{Kaper_1966,Reid_1988,Condon_1997})

\begin{equation}
\sigma_x \approx \frac{1}{2} \frac{\Delta_x}{\text{\snr}} . \label{eqn:GaussianFit}
\end{equation}

\noindent We find that maser spots with measured \snr$ \geq 3$ have uncertainties that are well-described by the above expression (see \autoref{app:DiskLikelihood}), so we use only such maser spots for the measurements in this paper.

\subsection{Measuring accelerations from monitoring spectra} \label{AccelerationFitting}

We measured accelerations using a time-dependent Gaussian decomposition of the maser spectrum.  For each of $N$ Gaussians, the free parameters are the amplitude $A$, the linewidth $\sigma$, the initial central velocity $v_0$ (referenced to a particular observing epoch), and its linear drift in time $a$ (i.e., the measured acceleration).  The model spectrum at each epoch $t$ (where $t=0$ corresponds to the reference epoch) is then obtained by summing each of the individual Gaussians,

\begin{equation}
S(v,t) = \sum_{i=1}^N A_i(t) \exp\left( - \frac{\left[ v - (v_{0,i} + a_i t) \right]^2}{2 \sigma_i^2} \right) , \label{eqn:ModelSystemicSpectrum}
\end{equation}

\noindent where the index $i$ indicates the values for the $i$th Gaussian.  The individual amplitudes are allowed to vary from one epoch to the next, while the line widths are held fixed.  The fitting was performed using a least-squares routine, choosing random initial guesses for each of the parameters.  Nine consecutive monitoring epochs were fit simultaneously, and the fitting procedure was repeated 100 times for each set of nine consecutive spectra (approximating the lifetime of an individual maser feature), with typical reduced-$\chi^2$ values between 1.2 and 1.7.  From each suite of 100 fits, the 10 best (i.e., smallest reduced-$\chi^2$) were selected and averaged to produce the final acceleration measurements.  See ``Method 2'' from \cite{2013ApJ...767..154R} for further details regarding the fitting code.

We applied this fitting technique separately to each group of maser features.  The systemic features were fit within the velocity range 6880--6975\,\kms, the blueshifted features within the range 6120--6405\,\kms, and the redshifted features within the range 7580--7780\,\kms.  The best-fit accelerations were binned as a function of $v_0$ to match the VLBI spectral binning, and an acceleration measurement was assigned to each channel as the $\chi^2$-weighted mean of the 10 best-fit accelerations within that channel.  Several of the high-velocity maser features detected in the VLBI map, particularly in the blueshifted complex, are too weak in individual monitoring spectra to obtain an acceleration measurement.  Nevertheless, it is useful to include these data points in the disk modeling, as they are still capable of providing model constraints (e.g., on the location and velocity of the dynamic center).  In past MCP works, we have assigned a nominal acceleration measurement (e.g., $0 \pm 1$\,\kmsyr) to such weak maser features; here, we opt for an alternative treatment that uses the ensemble of measurements to constrain the unmeasured accelerations within the context of a complete disk model.  See \autoref{app:DiskLikelihood} for details.

\begin{figure*}[t]
	\centering
		\includegraphics[width=1.00\textwidth]{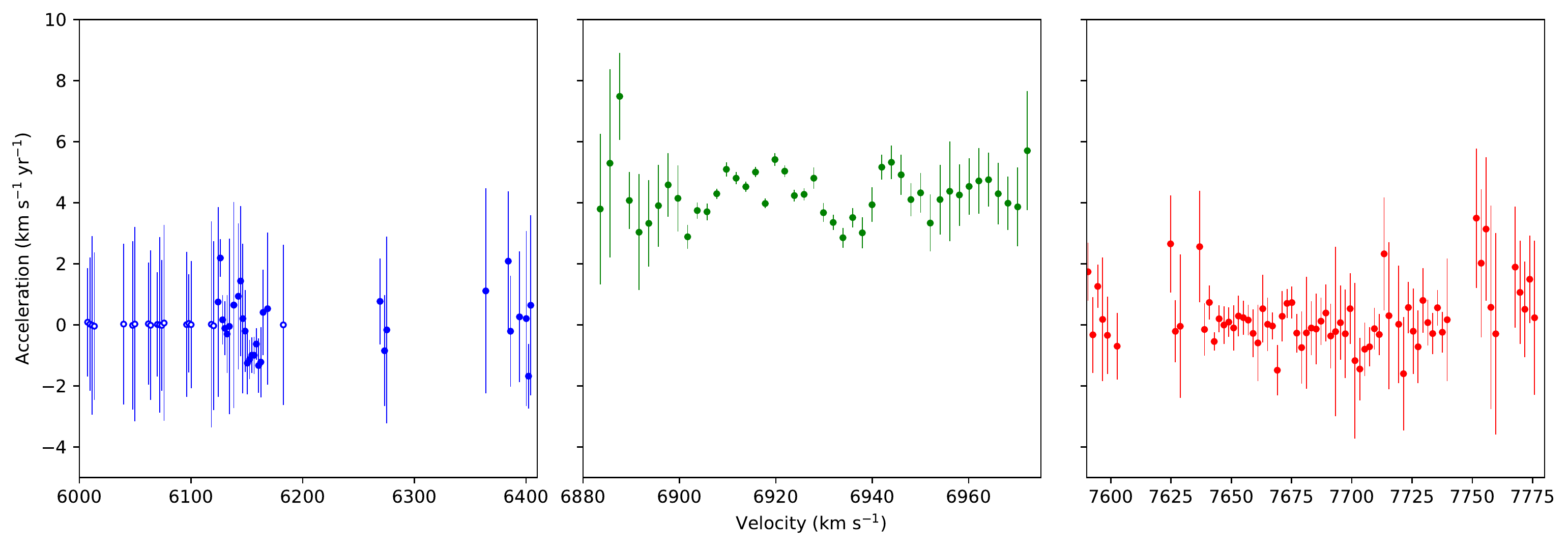}
	\caption{Acceleration measurements for the three sets of maser features, with data points colored by velocity complex (blue for blueshifted features, green for systemic features, and red for redshifted features).  The accelerations have been measured as described in \autoref{AccelerationFitting}.  Unmeasured accelerations (i.e., those that were fit by the model) are plotted using open circles.}
	\label{fig:accel_plot}
\end{figure*}

The accelerations and their uncertainties are shown in \autoref{fig:accel_plot}.  We can see that the systemic features share a roughly constant acceleration with a mean of 4.38\,\kmsyr and an RMS of 0.66\,\kmsyr, suggesting that they mostly reside in a thin annulus with little ($\sim$5\%) spread in orbital radius.  The redshifted features show a mean acceleration of $0.06$\,\kmsyr (RMS of 0.65\,\kmsyr), and the blueshifted features have a mean of $-0.35$\,\kmsyr (RMS of 1.11\,\kmsyr).  Both sets of high-velocity features have accelerations that are consistent with zero, as expected for masers located near the midline of the disk.

\section{Determining the Hubble constant} \label{sec:HubbleConstant}

To measure $H_0$, we fit a three-dimensional warped disk model to the $(x,y,v,a)$ measurements obtained for each velocity channel of the VLBI map.  The basic model is a thin, warped disk in which clouds of masing gas orbit a massive point source.  The model can include three parameters describing confocal elliptical instead of pure circular orbits, although for CGCG 074-064 the data did not support this addition (see \autoref{sec:FittingProcedure}).  The relative weightings of the heterogeneous $(x,y,v,a)$ data products are important and are discussed in detail in \autoref{app:DiskLikelihood}.

Our fitting procedure requires that we explore a moderately high-dimensional ($d > 300$, mostly nuisance parameters) parameter space subject to several strong correlations between model parameters (e.g., between $D$ and $M_{\text{BH}}$).  We perform the model fitting within a Bayesian framework, whereby the posterior distribution $\mathcal{P}(\boldsymbol{\Theta} | \textbf{D})$ of the model parameters $\boldsymbol{\Theta}$ conditioned on the data $\textbf{D}$ is given by Bayes' theorem,

\begin{equation}
\mathcal{P}(\boldsymbol{\Theta} | \textbf{D}) = \frac{\mathcal{L}(\boldsymbol{\Theta}) \pi(\boldsymbol{\Theta})}{\mathcal{Z}} .
\end{equation}

\noindent Here, $\mathcal{L}(\boldsymbol{\Theta}) = \mathcal{P}(\textbf{D} | \boldsymbol{\Theta})$ is the likelihood of the data conditioned on the model parameters, $\pi(\boldsymbol{\Theta})$ is the prior probability of the model parameters, and $\mathcal{Z} = \int \mathcal{L}(\boldsymbol{\Theta}) \pi(\boldsymbol{\Theta}) \text{d}\boldsymbol{\Theta}$ is the Bayesian evidence.

Past MCP papers have used a random walk Metropolis-Hastings (MH) Markov Chain Monte Carlo (MCMC) algorithm to sample the posterior (see \citealt{2013ApJ...767..154R} for a detailed description), but in this work we introduce instead a new algorithm utilizing the Hamiltonian Monte Carlo \citep[HMC;][]{Neal_2012} sampler implemented in PyMC3\footnote{\url{https://github.com/pymc-devs/pymc3}} \citep{Salvatier2016}.  HMC methods take advantage of the posterior geometry to efficiently explore the ``typical set'' (i.e., the region containing the bulk of the probability mass) even in complex and high-dimensional spaces; see \cite{2017arXiv170102434B} for a concise overview of HMC.  In addition to increased sampling efficiency, the primary improvement provided by the new disk-fitting code is the ability to fit for the ``error floor'' parameters as part of the model, thereby removing a source of systematic uncertainty that has limited the precision of previous MCP measurements.

In this section we focus on the disk-fitting procedure and resulting measurement of the Hubble constant.  A comprehensive description of the disk model is provided in \autoref{app:DiskModel}, and \autoref{app:DiskLikelihood} details the construction of the likelihood function.

\subsection{Fitting procedure and systematic error estimation} \label{sec:FittingProcedure}

The primary end results of our fitting procedure are point estimators -- namely the median value and some confidence interval around it -- for each of the model parameters.  To achieve some desired level of precision $p$ in such estimators requires ${\sim}1/p^2$ independent samples.  Adjacent samples in the MCMC chain are generally correlated, so we use the autocorrelation time, $\tau$, to determine our effective sample size.

The autocorrelation $A(t)$ for a chain is given by

\begin{equation}
A(t) = \frac{\sum\limits_{i} \left( \boldsymbol{\Theta}_i - \left\langle \boldsymbol{\Theta} \right\rangle \right)^{\top} \left( \boldsymbol{\Theta}_{i+t} - \left\langle \boldsymbol{\Theta} \right\rangle \right)}{\sum\limits_{i} \left| \boldsymbol{\Theta}_i - \left\langle \boldsymbol{\Theta} \right\rangle \right|^2} ,
\end{equation}

\noindent where $\boldsymbol{\Theta}_i$ is the parameter column vector at the $i$th step of the chain, $t$ is the lag, and angle brackets $\left\langle \right\rangle$ denote a sample average.  We define the integrated autocorrelation time $\tau$ to be the sum of $A(t)$ over all $t \geq 0$,

\begin{equation}
\tau = \sum_{t} A(t) ,
\end{equation}

\noindent such that the value of $\tau$ indicates roughly how many steps in the chain separate two independent samples \citep{Sokal_1997}.  The effective sample size is then given by the number of autocorrelation times contained in the chain,

\begin{equation}
N_{\text{eff}}= \frac{N}{\tau} .
\end{equation}

\noindent We aim to achieve 1\% precision in our parameter estimates, requiring an effective sample size of $N_{\text{eff}} \approx 10^4$.

The ``best-fit'' parameters we recover will depend on the specific choice of model used, and we don't know \textit{a priori} which model parameters will be well-constrained by the data and which will prove to be extraneous.  We have thus performed a series of model fits using different assumptions for which parameters are permitted to vary in the underlying models and for how strong of a prior we impose on the error floor parameters.  We use these different model specifications to assess the magnitude of the model-dependent systematics in our measurements.  The results from these different model-fitting runs indicate the following:

\begin{enumerate}
    \item The data are unable to place useful constraints on the inclination warp, and models that permit a warp in the inclination direction find posteriors for $\frac{di}{dr}$ that are consistent with zero.  The inclusion of an inclination warping parameter in the model shifts the posterior for $D$ by $\sim$1\%.
    \item The data are unable to place useful constraints on the eccentricity or periapsis angle (and associated warping parameter), and models that permit these parameters to vary find posteriors that are consistent with the masers being on purely circular orbits.  The inclusion of eccentricity parameters in the model shifts the posterior for $D$ by $\sim$3\%.
    \item The data are able to place useful constraints on the error floor parameters, and while the specific choice of prior -- i.e., uniform (\autoref{eqn:UniformPrior}) or Normal (\autoref{eqn:NormalPrior}) -- used for these parameters affects their recovered values, the prior choice does not significantly impact the posteriors of the other parameter values.  Choosing one or the other prior type shifts the posterior for $D$ by $\sim$1\%.
\end{enumerate}

\noindent We find typical autocorrelation times of $\tau \lesssim 10$ for models that assume no eccentricity, and of $\tau \lesssim 100$ for those that permit eccentric orbits.  To achieve our desired $N_{\text{eff}} = 10^4$, we thus run chains of length $N = 10^5$ for the former models and of length $N = 10^6$ for the latter.

Given the above findings, we choose for our fiducial fit a model that applies uniform priors to the error floor parameters, assumes that the disk has no warping in the inclination direction, and assumes that the maser orbits are perfectly circular.  For each parameter, we use the variance-weighted standard deviation of the values produced by the suite of different model fits to estimate the systematic uncertainty associated with our model selection (see \autoref{tab:DiskFittingResults}).

\subsection{Results from disk fitting} \label{sec:DiskFittingResults}

\begin{figure*}[t]
	\centering
		\includegraphics[width=1.00\textwidth]{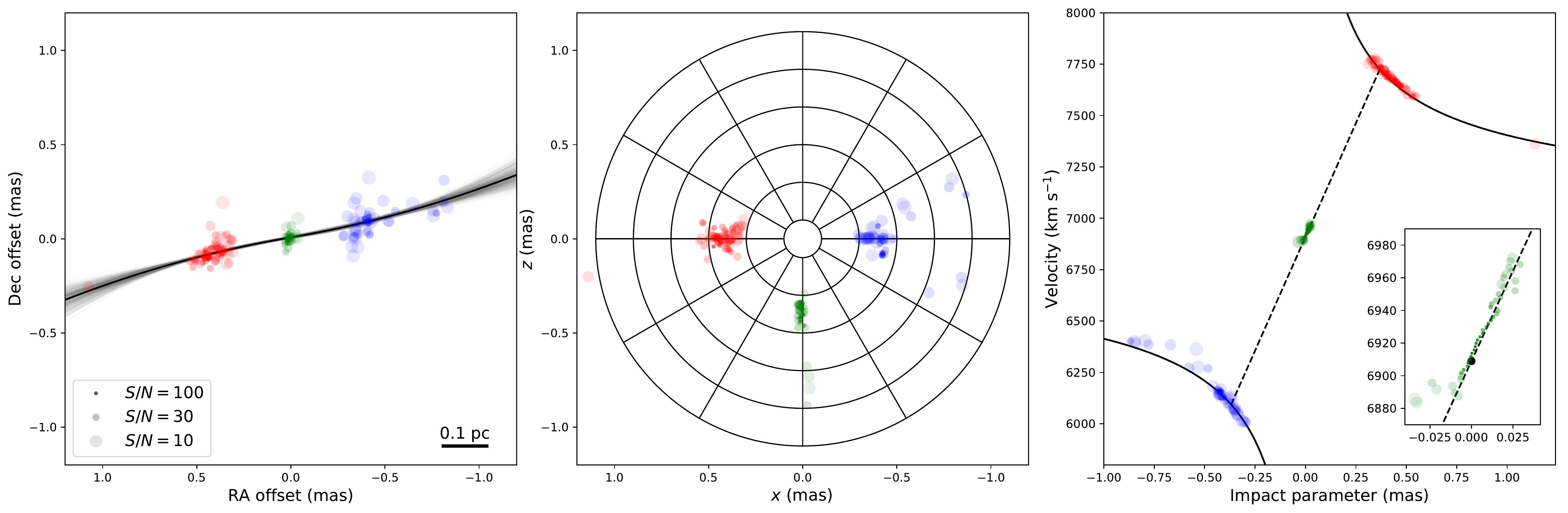}
	\caption{Map of the maser distribution in CGCG 074-064 atop our best-fit warped disk model as seen in the sky plane (left) and face-on (center), and the corresponding rotation curve (right).  The data points are colored by velocity group, with the red points corresponding to redshifted features, the blue points to blueshifted features, and the green points to systemic features.  The colors are darker for higher \snr, and the symbol sizes are proportional to $(\text{\snr})^{-1/2}$ (so that data points with larger uncertainties appear larger; see the legend in the left panel).  In all panels, the solid black lines trace the best-fit disk model, while in the left panel the light gray lines show the fits from 1000 different samplings of the posterior distribution.  In the right panel the dashed black line shows the average annulus for the systemic features (i.e., if the systemic features all originate from a thin ring at a single orbital radius, we would expect them to fall on or near this line), and the inset plot shows a zoom-in on the systemic features with the location of the SMBH marked as a black point.  The ``impact parameter'' is defined to be $r \sin(\phi)$ for every maser spot, with $r$ the orbital radius and $\phi$ the azimuthal angular position in the disk (see \autoref{app:DiskModel}).}
	\label{fig:disk_map_and_rotcurve}
\end{figure*}

We find that our disk model provides a good fit to the data, obtaining a $\chi^2$ value of 266.7 for 256 degrees of freedom. \autoref{tab:DiskFittingResults} lists the best-fit values and associated uncertainties for all modeled disk parameters, and \autoref{fig:correlation_diagram} shows the 1D and 2D posterior distributions for the same set of parameters.  \autoref{fig:disk_map_and_rotcurve} shows a map of the maser system as seen in the sky plane (left panel) and in the plane of the disk (central panel), overplotted on the best-fit disk model.  We can see from the map that the maser disk shows a modest warp in position angle, and from the model exploration described in \autoref{sec:FittingProcedure} we find that the disk is consistent with having zero warping in the inclination direction.

The right panel of \autoref{fig:disk_map_and_rotcurve} shows the maser rotation curve, which is consistent with the Keplerian behavior expected for material orbiting in a point-source potential.  We constrain the mass of the SMBH in CGCG 074-064 to be $2.42^{+0.22}_{-0.20} \times 10^7$\,M$_{\odot}$, comparable to other megamaser systems which typically have SMBH masses of $\sim$10$^7$\,M$_{\odot}$ \citep[see, e.g.,][]{2011ApJ...727...20K, 2017ApJ...834...52G}.  The innermost masers reside at orbital radii of $\sim$0.3~mas, corresponding to $\sim$0.12~pc (${\sim}5.5 \times 10^4$ Schwarzschild radii) at the best-fit angular-size distance of 87.6\,Mpc to CGCG 074-064.  From the right panel of \autoref{fig:disk_map_and_rotcurve} we can see that the maser velocities are well-fit by a Keplerian rotation curve even at these inner radii, implying a lower limit on the mass density of the enclosed object of $\rho \gtrsim 3 \times 10^9$~M$_{\odot}$~pc$^{-3}$.

\subsection{Measuring $H_0$} \label{sec:MeasuringH0}

Our disk modeling does not directly return a posterior distribution for $H_0$, but instead constrains both the angular-size distance to the maser disk ($D$) and the central SMBH redshift ($z_0$) separately.  If the SMBH is at rest with respect to the host galaxy, and if that galaxy has no peculiar motion with respect to the Hubble flow, then we can determine $H_0$ from these values using an expression adapted from \cite{1999astro.ph..5116H},

\begin{equation}
H_0 = \frac{c}{D \left( 1 + z_0 \right)} \int_0^{z_0} \frac{dz}{\sqrt{\Omega_m \left( 1 + z \right)^3 + \left( 1 - \Omega_m \right)}} , \label{eqn:H0Integral}
\end{equation}

\noindent which assumes a flat $\Lambda$CDM cosmology.  We use the matter density parameter value from \cite{Planck_2018}, namely $\Omega_m = 0.315$.  For CGCG 074-064, applying \autoref{eqn:H0Integral} results in a $\sim$2.7\% reduction in the value of $H_0$ compared to simply using $H_0 = cz _0/D$.

We account for the peculiar motion of CGCG 074-064 by replacing $z_0$ in \autoref{eqn:H0Integral} with the redshift $z_{\text{flow}}$ from the \textit{Cosmicflows-3} database \citep{Tully_2015}.  At our measured angular-size distance of $87.6$\,Mpc in the direction of CGCG 074-064, the expected recession velocity is $v_{\text{flow}} = c z_{\text{flow}} = 7308 \pm 150$\,\kms using the optical convention in the CMB reference frame \citep[][see also the online calculator\footnote{\url{http://edd.ifa.hawaii.edu/CF3calculator/}}]{Graziani_2019}, implying a line-of-sight peculiar velocity for the galaxy of ${\sim}136$\,\kms.

\begin{figure*}[t]
	\centering
		\includegraphics[width=1.00\textwidth]{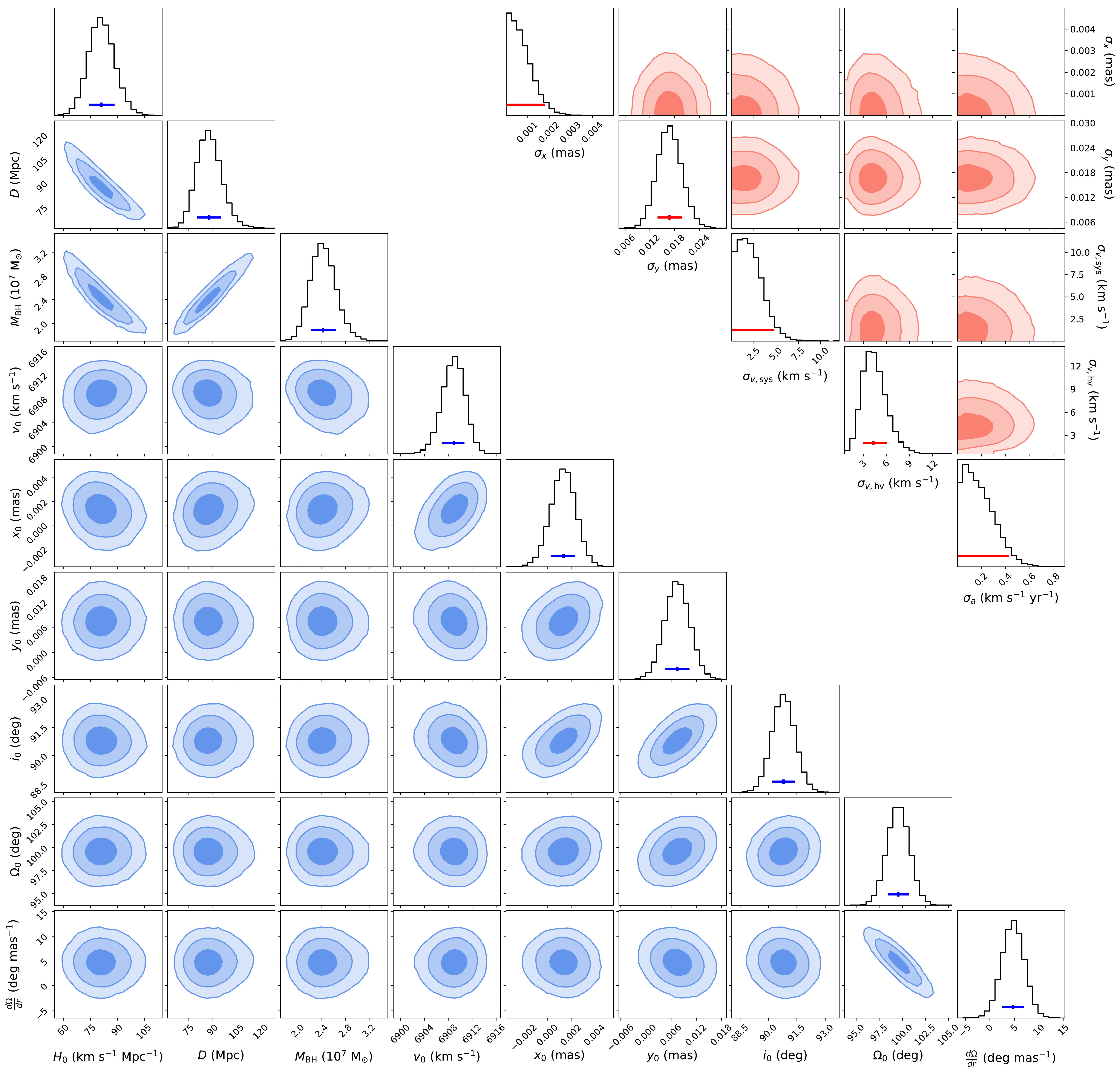}
	\caption{\textit{Lower left}: 1D posterior distributions (diagonal) and pairwise 2D posterior distributions (lower triangle) for $H_0$ (leftmost column) and the global parameters of the warped disk model fit to the maser system in CGCG 074-064.  The contours enclose 50\%, 90\%, and 99\% of the posterior probability.  The blue horizontal bars below each 1D histogram show the range from 16th to 84th percentile, with the 50th percentile point marked.  \textit{Upper right}: 1D marginalized posterior distributions (diagonal) and pairwise marginalized 2D distributions (upper triangle) for the error floor parameters from the same warped disk model fit.  The contours again enclose 50\%, 90\%, and 99\% of the posterior probability.  For the $\sigma_y$ and $\sigma_{v,\text{hv}}$ parameters, the horizontal bars below each 1D histogram show the range from 16th to 84th percentile, with the 50th percentile point marked.  For the $\sigma_x$, $\sigma_{v,\text{sys}}$, and $\sigma_a$ parameters, the horizontal bars indicate the 95\% upper limit region.}
	\label{fig:correlation_diagram}
\end{figure*}

Our resulting constraint on the Hubble constant is $H_0 = 81.0^{+7.4}_{-6.9}$\,\kmsmpc, with the posterior distribution shown in \autoref{fig:correlation_diagram}.

\begin{deluxetable*}{LcCCCC}
\tablecolumns{6}
\tablewidth{0pt}
\tablecaption{Disk fitting results for CGCG 074-064\label{tab:DiskFittingResults}}
\tablehead{\colhead{Parameter}	&	\colhead{Units}	&	\colhead{Prior}	&	\colhead{Posterior median} & \colhead{Statistical uncertainty} & \colhead{Systematic uncertainty}}
\startdata
D &	Mpc	& \mathcal{U}(10,150)\tablenotemark{a} &	87.6 & (+7.9,-7.2) & 1.5 \\
M_{\text{BH}} &	$10^7$ M$_{\odot}$ & \mathcal{U}(0.1,10.0) & 2.42 & (+0.22,-0.20) & 0.05 \\
v_0\tablenotemark{b} &	\kms & \mathcal{U}(6500,7500) &	6908.9 & (+1.8, -1.9) & 1.7 \\
x_0 & mas & \mathcal{U}(-0.5,0.5) & 0.0013 & (+0.0010,-0.0011) & 0.0010 \\
y_0 & mas & \mathcal{U}(-0.5,0.5) & 0.0075 & (+0.0029,-0.0029) & 0.0008 \\
i_0 & degree & \mathcal{U}(70,110) & 90.8 & (+0.6, -0.6) & 1.1 \\
\Omega_0 & degree & \mathcal{U}(0,180) & 99.6 & (+1.2, -1.2) & 0.6 \\
\frac{d\Omega}{dr} & degree\,mas$^{-1}$ & \mathcal{U}(-100,100) & 4.7 & (+2.2, -2.2) & 1.4 \\
\midrule
\sigma_x & mas & \mathcal{U}(0.0,0.1) & < 0.0018 & \ldots & 0.0001 \\
\sigma_y & mas & \mathcal{U}(0.0,0.1) & 0.017 & (+0.003, -0.003) & 0.0009 \\
\sigma_{v,\text{sys}} & \kms & \mathcal{U}(0,20) & < 4.8 & \ldots & 0.1 \\
\sigma_{v,\text{hv}} & \kms & \mathcal{U}(0,20) & 4.3 & (+1.7, -1.4) & 0.5 \\
\sigma_a & \kmsyr & \mathcal{U}(0,20) & < 0.43 & \ldots & 0.03 \\
\midrule
H_0  & \kmsmpc  & \ldots & 81.0 & (+7.4, -6.9) & 1.4 \\
\enddata
\tablecomments{\textit{Top}: Fitting results for the global parameters describing the maser disk, marginalized over all other parameters.  Here, $D$ is the angular-size distance to the galaxy, $M_{\text{BH}}$ is the mass of the SMBH, $v_0$ is the line-of-sight velocity of the SMBH, $(x_0,y_0)$ is the coordinate location of the SMBH (relative to the phase center defined by the strongest systemic features; see \autoref{VLBIObs}), $i_0$ is the inclination angle of the disk at $r=0$, $\Omega_0$ is the position angle of the disk at $r=0$, and $\frac{d\Omega}{dr}$ is the first-order position angle warping parameter.  For the statistical uncertainty we quote 1$\sigma$ confidence intervals from the posterior, while for the systematic uncertainty we quote the weighted standard deviation across all model fits.  \textit{Middle}: Fitting results for the error floor parameters; $\sigma_x$ is the $x$-position error floor, $\sigma_y$ is the $y$-position error floor, $\sigma_{v,\text{sys}}$ is the error floor for the systemic feature velocities, $\sigma_{v,\text{hv}}$ is the error floor for the high-velocity feature velocities, and $\sigma_a$ is the acceleration error floor.  For $\sigma_x$, $\sigma_{v,\text{sys}}$, and $\sigma_a$, 95\% upper limits are quoted rather than posterior medians.  \textit{Bottom}: Hubble constant measurement derived from the disk fit.}
\tablenotetext{a}{$\mathcal{U}(a,b)$ denotes a uniform distribution on the range $[a,b]$; see \autoref{eqn:UniformPrior}.}
\tablenotetext{b}{We directly model $z_0$ in the CMB frame (see \autoref{app:DiskModel}), which we have converted in this table to $v_0$ (optical convention) in the barycentric frame.  The conversion is $v_0 = c z_0 - 263.3$\,\kms.}
\end{deluxetable*}

\section{Discussion} \label{Discussion}

\subsection{Intermediate-velocity masers}

\autoref{fig:GBT_spectrum} shows an average over nearly two years of maser spectra taken with the GBT towards CGCG 074-064, and it represents the most sensitive spectrum we have of this source.  A number of faint maser lines are observed in this deep GBT spectrum that do not appear in the VLBI spectrum, either for reasons of insufficient sensitivity or because of a finite feature lifetime, and several of these ``intermediate-velocity'' lines have apparent orbital velocities that are considerably lower than those of the maser spots mapped with VLBI.

The highest-velocity blueshifted feature in the VLBI map is located at $\sim$6404\,\kms, while the highest-velocity blueshifted feature in the deep GBT spectrum is located at $\sim$6589\,\kms.  Its corresponding orbital velocity of $\sim$322\,\kms would put this maser spot at an orbital radius of ${\sim}0.89$\,pc (${\sim}2.2$\,mas) -- roughly twice as far away from the SMBH as the most distant mapped maser spot.  A similar story holds for the redshifted features, with the lowest-velocity redshifted feature in the deep GBT spectrum located at $\sim$7329\,\kms, corresponding to an orbital velocity of $\sim$419\,\kms and an orbital radius of ${\sim}0.56$\,pc (${\sim}1.4$\,mas).

One intermediate-velocity redshifted feature (located at $\sim$7363\,\kms) is detected in the VLBI map at an apparent radius of $\sim$1.1\,mas (see \autoref{fig:map_and_continuum}), providing direct evidence that the population of intermediate-velocity masers continues to follow the rotation curve traced by the high-velocity features out to at least $\sim$0.44\,pc.  It thus appears that the region of the accretion disk capable of supporting maser activity likely extends well beyond what we have mapped.  Next-generation radio facilities can be expected to regularly detect these faint features and thereby provide significantly more comprehensive coverage of the accretion disk geometry.

\subsection{Maser disk scale height}

The $y$-position error floors ($\sigma_y \approx 17$\,$\mu$as) are at least an order of magnitude larger than their $x$-position counterparts ($\sigma_x \lesssim 2$\,$\mu$as, consistent with zero), though the beam elongation is only a factor of $\sim$3 greater in the $y$-direction than in the $x$-direction.  If we assume that the measurement error contribution to $\sigma_y$ scales with $\sigma_x$ in the same manner as the beam, then we find $\sigma_{\text{excess}} = \sqrt{\sigma_y^2 - (3 \sigma_x)^2} \approx 16$\,$\mu$as of excess scatter in the $y$-direction.  The fortuitous East-West orientation of the maser disk on the sky permits us to use this excess scatter to set an upper limit on the maser disk scale height of $h \lesssim \sigma_{\text{excess}}$, or $h \lesssim 1300$\,AU at the distance to CGCG 074-064.  This scale height corresponds to a disk aspect ratio of $r/h \gtrsim 100$, which is firmly in the thin disk regime (i.e., $r \gg h$).

For a thin disk in hydrostatic equilibrium, the disk aspect ratio is equal to the Mach number of the orbiting material \citep{2002apa..book.....F}.  The orbital velocities of the masers in CGCG 074-064 span from $\sim$400--900\,\kms, implying local sound speeds of $c_s \lesssim 4$--9\,\kms.  These limits are consistent with the $c_s = 1.5$\,\kms measured by \cite{2007ApJ...659.1040A} for the accretion disk around the SMBH in the galaxy NGC 4258 \citep[see also][]{1995PNAS...9211427M}.

\subsection{VLBI continuum emission}

The continuum source detected in our VLBI data (see \autoref{fig:map_and_continuum}) shows a perpendicular offset from the maser disk, reminiscent of the continuum structure seen towards the disk in NGC 4258 \citep{1997ApJ...475L..17H} and suggesting a jet origin.  With a peak surface brightness of $46 \pm 9.5$~$\mu$Jy~beam$^{-1}$, the brightness temperature of the continuum source is at least $1.6 \times 10^6$~K.

The measured disk inclination angle indicates that we are seeing the central region of the disk almost perfectly edge-on ($i_0 \approx 91$\,degrees; see \autoref{tab:DiskFittingResults}), implying that if relativistic beaming is the cause of the apparent one-sidedness of the (presumably intrinsically symmetric) jet, then the jet axis must be misaligned with the angular momentum vector of the disk on $\sim$0.1\,pc scales.  We can estimate the required magnitude of misalignment using an approach similar to that used in \cite{1989ApJ...336..112R}, by assuming that the jet in CGCG 074-064 is intrinsically symmetric (in both velocity and emitted spectral power) and that the northern (observed) component is approaching us while the southern (unobserved) component is receding from us.

For a jet of material moving at constant speed $\beta \equiv v/c$, the observed flux density $S$ is related to the flux density $S_0$ that would be observed if the source were at rest by \citep{1967MNRAS.136..123R}

\begin{equation}
S = \frac{S_0}{\left( \gamma \left[ 1 - \beta \cos(\theta) \right] \right)^{3+\alpha}} , \label{eqn:RelativisticFluxDensity}
\end{equation}

\noindent where $\gamma = \left( 1 - \beta^2 \right)^{-1/2}$ is the Lorentz factor, $\alpha$ is the spectral index (such that $S \propto \nu^{-\alpha}$), and $\theta$ is the angle between the velocity vector and the line of sight (such that $\theta = 0$ corresponds to material moving directly towards us).  The ratio of the observed flux densities of the northern ($S_N$) and southern ($S_S$) components will then be

\begin{equation}
\frac{S_N}{S_S} = \left( \frac{1 + \beta \cos(\theta)}{1 - \beta \cos(\theta)} \right)^{3+\alpha} \geq \frac{46}{19} , \label{eqn:RelativisticFluxDensityRatio}
\end{equation}

\noindent where we measure the flux density of the northern (approaching) component to be $46$~$\mu$Jy, and we have used $19$\,$\mu$Jy (i.e., twice the RMS in the continuum image) as an upper limit for the flux density of the (unobserved) southern component.

If we assume that the jet is aligned with the angular momentum vector of the disk, then $\theta$ would be equal to $89$\,degrees and the inequality in \autoref{eqn:RelativisticFluxDensityRatio} could not be satisfied for any value of $\beta$.  In general, the minimum degree of disk-jet misalignment needed to satisfy \autoref{eqn:RelativisticFluxDensityRatio} for a given value of $\theta$ will occur when $\beta=1$, so we can impose this value to determine a lower limit on the misalignment.  For a typical spectral index of $\alpha = 0.7$ \citep{1998AJ....115.1693C, 2017A&A...605A..84K}, we find that the disk-jet misalignment must be at least ${\sim}7^{\circ}$; even for a spectral index as steep as $\alpha = 2$, the misalignment can be no smaller than ${\sim}5^{\circ}$.  Using a more conservative $3{\sigma}$ upper limit for the unmeasured southern jet component yields misalignment angles of at least ${\sim}5^{\circ}$ and ${\sim}3^{\circ}$ for $\alpha = 0.7$ and $\alpha = 2$, respectively.

An alternative explanation for the lack of an observed southern jet component could be that our line of sight to that component passes through the maser disk, while emission from the northern jet component reaches us unimpeded.  X-ray irradiation from the central AGN may produce layers of hot (${\sim}10^4$\,K) ionized material above and below the molecular disk, providing an absorption opportunity for emission that must pass through the disk along the line of sight \citep{1995ApJ...447L..17N}.  \cite{1996ApJ...468L..17H} estimate the free-free optical depth in this layer to be $\sim$2--3 at a frequency of 22~GHz for the disk in NGC 4258; a similar level of absorption would be sufficient to explain the asymmetric flux densities of the two jet components in CGCG 074-064.

If the nuclear continuum provides the seed photons for the systemic maser complex, as suggested by \cite{1995Natur.373..127M} for NGC 4258, then we can estimate the maser gain from the observed strength of the systemic features.  In the absence of free-free attenuation, a $\sim$45~$\mu$Jy continuum would require an amplification of $3 \times 10^3$ to power the $\sim$150~mJy systemic masers (corresponding to an optical depth of $\tau \approx -8$).  This level of amplification is similar to that inferred for the systemic features in NGC 4258 \citep{1997ApJ...475L..17H}.

\subsection{Variability of the maser features} \label{MaserVariability}

In both our GBT and VLA monitoring observations of CGCG 074-064, we found that the strongest systemic features (i.e., those that could be identified in individual scans) often show substantial ($\sim$50\%) variability on timescales of $\sim$tens of minutes.  In one case -- that of the 6915\,\kms line during the 2016 October GBT observation -- the flux density increased by a factor of $\sim$3 over the course of half an hour from 100~mJy to 300~mJy; if intrinsic to the maser system, this behavior would correspond to an increase of $\sim$60~$L_{\odot}$ in isotropic luminosity across a region no larger than $\sim$3~AU in size (as determined by light-travel time).  The magnitude of this variability, and the fact that it is uncorrelated between different maser features, indicates that observational effects (e.g., fluctuations in antenna gain or atmospheric opacity) are unlikely to be the cause.

Such rapid variability has been seen before in at least three other H$_2$O megamaser systems -- Circinus \citep{1997ApJ...474L.103G}, NGC 3079 \citep{2007ApJ...656..198V}, and ESO 558-G009 \citep{2015ApJ...810...65P} -- and in the IC 10 kilomaser system \citep{1994ApJ...422..586A}.  In all previous cases, interstellar scintillation (ISS) has been the preferred explanation for the observed variability.  However, the number of megamaser systems that are now known to display apparently ISS-induced variability, and the strength of the variability in these systems ($\sim$tens of percent or greater), appears to be largely unexpected.  Furthermore, it is not clear that ISS through a foreground Galactic scattering screen can satisfactorily account for the observed properties of this variability across all of these systems.  Some unresolved issues include:

\begin{enumerate}
    \item The number of megamaser galaxies that exhibit this class of variability is too large, and their on-sky locations are too removed from the Galactic plane.  The MASIV VLA survey \citep{2003AJ....126.1699L} found that only a tiny fraction ($\lesssim$1\%) of compact extragalactic radio continuum sources show strong ($\gtrsim$10\%) and rapid (timescales of $\sim$several hours) ISS-induced variability at an observing frequency of 5~GHz.  They also found the expected correlation between variability amplitude and line-of-sight emission measure from the Galactic ionized medium, resulting in greater variability being seen at lower absolute Galactic latitudes \citep{2008ApJ...689..108L}.  The presence of strong and rapid variability in $\gtrsim$10\% of all disk megamaser systems, and the lack of an obvious correlation with Galactic latitude (e.g., CGCG 074-064 is located at a Galactic latitude of $+$65 degrees), is then difficult to explain.
    \item The magnitude of the variability is too large.  For the weak scattering regime expected at high Galactic latitudes, the $\sim$tens of percent variability seen towards the MASIV targets at 5~GHz should be even smaller at the substantially higher 22~GHz observing frequency for the maser systems.
    \item The variability timescale is too short, and it is too consistent across different megamaser-hosting galaxies.  For ISS, the characteristic variability timescale is set by the transverse velocity of the scattering screen and the size of either the ``scintle'' (if the phase-coherent region of the scattering medium has a larger angular size than the source) or the source (if the source is larger in angular size than the scintle).  In the weak scattering limit, the scintle size goes as $\nu^{-1/2}$ \citep{1992RSPTA.341..151N}, so all else being equal we would in general expect only a factor of $\sim$2 shorter variability timescales at 22~GHz than what is seen at 5~GHz if the scintle sets the relevant size scale.  If instead the angular size of the source sets the relevant timescale then the magnitude of the variability would be decreased by a dilution factor roughly equal to the ratio of the source area to the scintle area, an expectation that is inconsistent with the observed (strong) variability.  Furthermore, the timescales for all scintillating maser galaxies are of the same order of magnitude while their distances differ by more than a factor of 20; if the maser spot size were setting the variability timescale, then we would expect the timescale to decrease inversely with the distance to the maser galaxy.
\end{enumerate}

Currently, we do not understand the cause of the variability seen towards CGCG 074-064.  It is perhaps the case that a scattering screen within the megamaser-hosting galaxy itself can explain the observed flux modulation, but we leave an investigation of this question to future work.

\section{Conclusions} \label{Conclusions}

We have presented a geometric distance measurement to the galaxy CGCG 074-064 of $87.6^{+7.9}_{-7.2}$\,(stat.)\,$\pm 1.5$\,(sys.)\,Mpc, made using the megamaser technique as part of the MCP.  The strength (typical flux density $>$200~mJy) and orderly accelerations (nearly constant at $4.4$\,\kmsyr across the entire systemic velocity complex) of the systemic features in this system have enabled a precise distance measurement with an uncertainty of only $\sim$9\%.  Our 3D warped disk modeling also constrains the mass of the SMBH in CGCG 074-064 to be $2.42^{+0.22}_{-0.20}$\,(stat.)\,$\pm 0.05$\,(sys.)\,$\times 10^7$\,M$_{\odot}$.  We have combined our angular-size distance measurement with a group recession velocity from \textit{Cosmicflows-3} to determine a value for the Hubble constant of $H_0 = 81.0^{+7,4}_{-6.9}$\,\kmsmpc, with a systematic error of $1.4$\,\kmsmpc.

Our VLBI observations of the maser system in CGCG 074-064 have also revealed a weak ($46 \pm 9.5$~$\mu$Jy~beam$^{-1}$), marginally-resolved continuum source that appears to originate from a nuclear jet.  The one-sided nature of this jet emission and its strength relative to the maser emission are both reminiscent of what has been previously seen in NGC 4258.  In addition, our spectral monitoring observations have revealed that the systemic maser features in CGCG 074-064 are highly variable, with flux densities changing by as much as a factor of 3 on timescales of tens of minutes.  Interstellar scintillation has been the preferred explanation for such variability in other megamaser systems, but we note unresolved issues with this explanation that will need to be addressed in future work.

\acknowledgments

We would like to honor the memory of our colleague, mentor, and friend Fred Lo.  We thank Adam Riess, Dan Scolnic, and Brent Tully for helpful discussions about peculiar velocities.  Support for this work was provided by the NSF through the Grote Reber Fellowship Program administered by Associated Universities, Inc./National Radio Astronomy Observatory.  The National Radio Astronomy Observatory is a facility of the National Science Foundation operated under cooperative agreement by Associated Universities, Inc.  This work made use of the Swinburne University of Technology software correlator, developed as part of the Australian Major National Research Facilities Programme and operated under license.  This research has made use of the NASA/IPAC Extragalactic Database (NED), which is operated by the Jet Propulsion Laboratory, California Institute of Technology, under contract with the National Aeronautics and Space Administration.  This work was supported in part by the Black Hole Initiative at Harvard University, which is funded by grants from the John Templeton Foundation and the Gordon and Betty Moore Foundation to Harvard University.

\vspace{5mm}
\facilities{GBT, VLA, VLBA, HSA}
\software{AIPS, CASA, GBTIDL, PyMC3 \citep{Salvatier2016}}

\clearpage
\appendix

\section{Disk model} \label{app:DiskModel}

Our disk model is very similar to that used by \cite{2013ApJ...767..154R} and \cite{2013ApJ...775...13H}.  We include global parameters describing the angular-size distance $D$ to the SMBH, the SMBH mass $M_{\text{BH}}$, the SMBH redshift $z_0$, and the on-sky coordinates of the SMBH $(x_0,y_0)$.  We also include several global parameters describing the warped geometry of the disk, which is parameterized by an inclination angle $i(r)$, a position angle $\Omega(r)$, and a periapsis $\omega(r)$ that vary as a function of orbital radius as

\begin{equation}
i(r) = i_0 + \frac{di}{dr} r , \label{eqn:IncWarp}
\end{equation}

\begin{equation}
\Omega(r) = \Omega_0 + \frac{d\Omega}{dr} r , \label{eqn:PosWarp}
\end{equation}

\begin{equation}
\omega(r) = \omega_0 + \frac{d\omega}{dr} r . \label{eqn:PeriWarp}
\end{equation}

\noindent The modeled geometric parameters are then $i_0$, $\frac{di}{dr}$, $\Omega_0$, $\frac{d\Omega}{dr}$, $\omega_0$, and $\frac{d\omega}{dr}$.

Each maser spot is assigned a location $(r,\phi)$ within the disk, where $r$ is the spherical radius measured from the BH and $\phi$ is the azimuthal angle measured from the line of sight (oriented such that the systemic features are located at $\phi \approx 0^{\circ}$ and the redshifted features are located at $\phi \approx 90^{\circ}$).  The sky-plane position of the maser spot is denoted $(x,y)$, with the $x$-axis aligned with right ascension (so that positive points to the east) and the $y$-axis aligned with declination (so that positive points to the north).  The $z$-axis is then directed along the line of sight, so that positive $z$ points away from us.  The inclination angle $i$ is defined to be the angle that the disk normal makes with respect to the line of sight (so that $90^{\circ}$ corresponds to perfectly edge-on).  The position angle $\Omega$ is then defined to be the angle that the receding portion of the disk midplane makes east of north (i.e., clockwise down from the $y$-axis).  Note that both $i$ and $\Omega$ are functions of $r$.

We can transform from the disk frame to the sky frame by rotating first by $i$ about the $x$-axis, then by $\Omega$ about the $z$-axis.  This transformation can be expressed as a product of two rotation matrices,

\begin{equation}
\left( \begin{array}{c}
x \\
y \\
z
\end{array} \right) = \left( \begin{array}{ccc}
\sin(\Omega) & - \cos(\Omega) & 0 \\
\cos(\Omega) & \sin(\Omega) & 0 \\
0 & 0 & 1
\end{array} \right) \left( \begin{array}{ccc}
1 & 0 & 0 \\
0 & \sin(i) & - \cos(i) \\
0 & \cos(i) & \sin(i)
\end{array} \right) \left( \begin{array}{c}
r \sin(\phi) \\
0 \\
- r \cos(\phi)
\end{array} \right) ,
\end{equation}

\noindent where we have for simplicity used a pre-rotation disk orientation of $i = \Omega = 90^{\circ}$.  After accounting for the location of the BH itself, we obtain the sky frame coordinates of the maser spot to be

\begin{subequations}
\begin{eqnarray}
x & = & x_0 + r \big[ \sin(\phi) \sin(\Omega) - \cos(\phi) \cos(\Omega) \cos(i) \big] , \\
y & = & y_0 + r \big[ \sin(\phi) \cos(\Omega) + \cos(\phi) \sin(\Omega) \cos(i) \big] , \\
z & = & - r \cos(\phi) \sin(i) .
\end{eqnarray}
\end{subequations}

\noindent (Note that the $z$ coordinate of the BH is fixed at $z=0$ by our choice of coordinate system.)  We can similarly express the sky frame components of the maser spot's acceleration as

\begin{subequations}
\begin{eqnarray}
a_x & = & a \big[- \sin(\phi) \sin(\Omega) + \cos(\phi) \cos(\Omega) \cos(i) \big] , \\
a_y & = & a \big[ - \sin(\phi) \cos(\Omega) - \cos(\phi) \sin(\Omega) \cos(i) \big] , \\
a_z & = & a \cos(\phi) \sin(i) ,
\end{eqnarray}
\end{subequations}

\noindent with

\begin{equation}
a(r) = \frac{G M_{\text{BH}}}{r^2 D^2} .
\end{equation}

\noindent Here, we've converted $r$ from angular units to physical ones using the angular-size distance to the SMBH, $D$.

The maser spot's velocity vector depends on the eccentricity $e$ and angle of periapsis $\omega$ of the orbit, and on the true anomaly $\psi = \phi - \omega$ of the spot.  For a maser spot situated at a particular $(r,\psi)$, the radial and azimuthal components of its velocity vector are

\begin{equation}
v_r = v(r) \frac{e \sin(\psi)}{\sqrt{1 + e \cos(\psi)}}
\end{equation}

\noindent and

\begin{equation}
v_{\phi} = v(r) \sqrt{1 + e \cos(\psi)} ,
\end{equation}

\noindent respectively, with

\begin{equation}
v(r) = \sqrt{\frac{G M_{\text{BH}}}{r D}} .
\end{equation}

\noindent Transforming to the sky coordinates, we obtain

\begin{subequations}
\begin{eqnarray}
v_x & = & v_{\phi} \big[ \cos(\phi) \sin(\Omega) + \sin(\phi) \cos(\Omega) \cos(i) \big] + v_r \big[ \sin(\phi) \sin(\Omega) - \cos(\phi) \cos(\Omega) \cos(i) \big] , \\
v_y & = & v_{\phi} \big[ \cos(\phi) \cos(\Omega) - \sin(\phi) \sin(\Omega) \cos(i) \big] + v_r \big[ \sin(\phi) \cos(\Omega) + \cos(\phi) \sin(\Omega) \cos(i) \big] , \\
v_z & = & v_{\phi} \sin(\phi) \sin(i) - v_r \cos(\phi) \sin(i) .
\end{eqnarray}
\end{subequations}

The redshift imparted by the relativistic Doppler effect is given by \citep{1986rpa..book.....R}

\begin{equation}
1 + z_D = \gamma \left( 1 - \frac{v}{c} \cos(\theta) \right) ,
\end{equation}

\noindent where $\gamma = \left( 1 - \frac{v^2}{c^2} \right)^{-1/2}$ is the Lorentz factor and $\theta$ is the angle between the velocity vector and the line of sight.  We can obtain $\cos(\theta)$ by taking the dot product between $\vec{v}/v$ and $-\hat{z}$, which results in

\begin{equation}
1 + z_D = \gamma \left( 1 + \frac{1}{c} \big[ v_{\phi} \sin(\phi) \sin(i) - v_r \cos(\phi) \sin(i) \big] \right) .
\end{equation}

In a Schwarzschild spacetime, the gravitational redshift $z_g$ of a photon emitted at radius $r$ and received at infinity is given by \citep{2009fcgr.book.....S}

\begin{equation}
1 + z_g = \left( 1 - \frac{R_s}{r D} \right)^{-1/2} ,
\end{equation}

\noindent where $R_s = 2 G M_{\text{BH}} / c^2$ is the Schwarzschild radius for the SMBH.

The observed redshift of the maser spot, $z$, will then be given by the product of both the Doppler and gravitational effects with the SMBH redshift, $z_0$:

\begin{equation}
1 + z = \left( 1 + z_D \right) \left( 1 + z_g \right) \left( 1 + z_0 \right) .
\end{equation}

\noindent Here, $z_0$ is the redshift measured in the CMB frame.  In this work we use the optical convention for all velocities, so the observed velocities are related to the redshift by simply

\begin{equation}
v_{\text{obs}} = c z .
\end{equation}

\section{Constructing the likelihood function} \label{app:DiskLikelihood}

For each data point $k$, we have a measurement of its on-sky position $(x_k,y_k)$, its line-of-sight velocity $v_k$, and its line-of-sight acceleration $a_k$.  Each of these measurements is independent of the others, and each is treated differently in the likelihood function.

\begin{figure*}[t]
	\centering
		\includegraphics[width=1.00\textwidth]{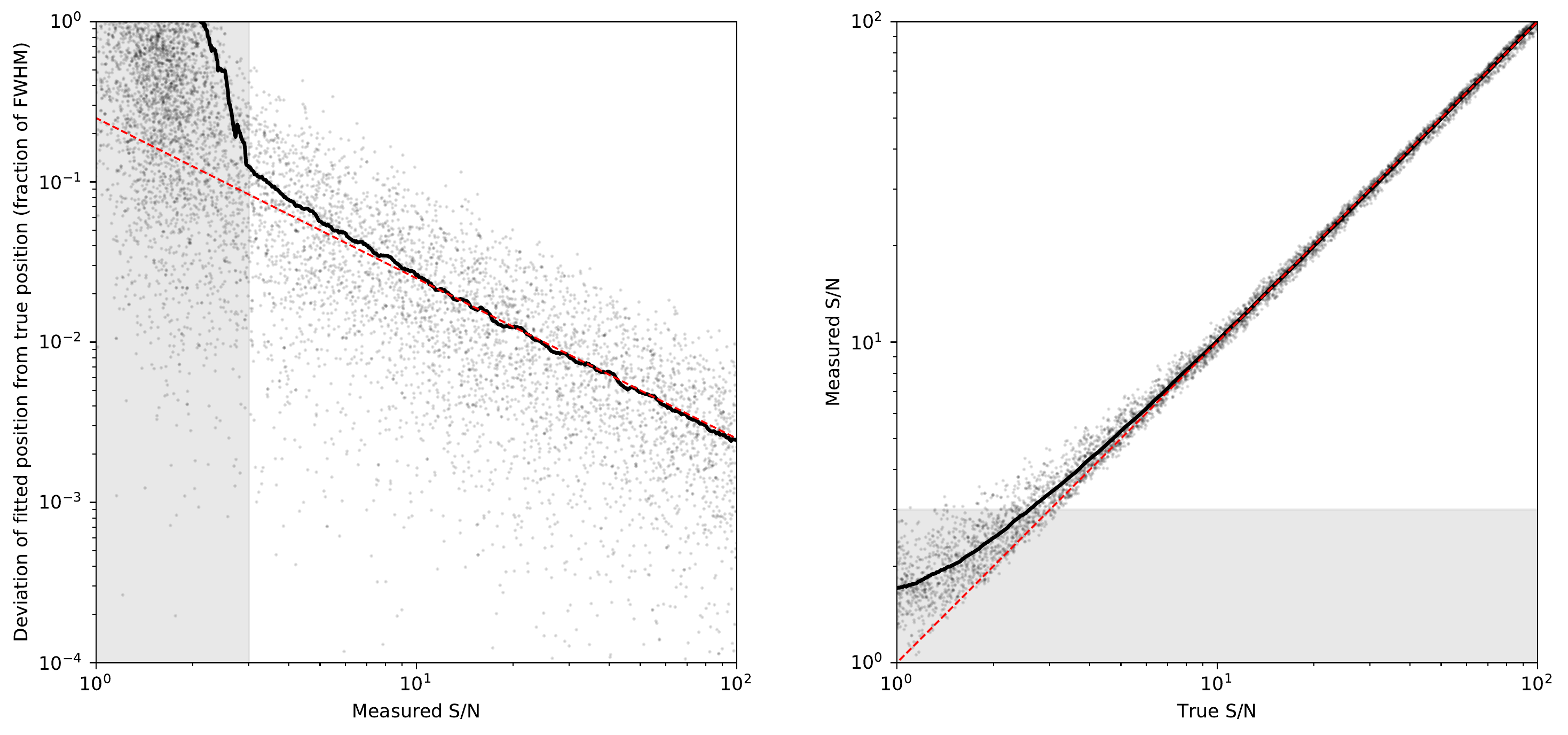}
	\caption{\textit{Left}: Centroid position measurement offsets versus \snr for a set of $10^5$ simulated point-source images.  The offsets are plotted as the absolute deviation between the measured position and the true position along a single axis (arbitrarily chosen to be the beam major axis), expressed as a fraction of the beam FWHM along that axis.  A running average every $10^3$ points is plotted as a solid black curve, and the theoretical noise limit given by \autoref{eqn:GaussianFit} is plotted as a dotted red curve.  We can see that the measurements adhere to the theoretical limit only for measured \snr values that are $\gtrsim$3.  \textit{Right}:  Measured \snr versus input \snr for the same set of $10^5$ simulated images.  For a real image we do not have access to the ``true'' \snr, so we would like to use only those for which the measured \snr matches it well.  We can see that our adopted $\text{\snr} \geq 3$ threshold, chosen so that the positional uncertainties match those given by \autoref{eqn:GaussianFit}, also allows us to exclude points for which our estimate of the \snr is unreliable.  As in the left panel, the solid black curve is a running average (every $10^3$ points), and the red dotted curve is the theoretical curve (in this case it is simply the $y=x$ line).  The gray shaded region in each plot indicates measurements that we would not have included in our final VLBI map (i.e., they fall below our measured \snr threshold).}
	\label{fig:snr_positional_accuracy}
\end{figure*}

The uncertainties $(\sigma_{x,k},\sigma_{y,k})$ in our position measurements are related to the beam dimensions by \autoref{eqn:GaussianFit}.  Though this expression was originally derived for an image with uncorrelated noise from pixel to pixel, we expect an analogous expression to hold in the case of oversampled data (for which noise will be correlated for pixels within a resolution element of one another).  To test this expectation, we fit a suite of $10^5$ mock images containing point sources and correlated noise corresponding to a known \snr (see \autoref{fig:snr_positional_accuracy}).  We found that for measured \snr values greater than $\sim$3, the uncertainties in the centroid coordinates are well-described by \autoref{eqn:GaussianFit}.  For our final VLBI map (see \autoref{fig:map_and_continuum}) we thus retain only those maser spots with measured $\text{\snr} \geq 3$, to which we assign positional uncertainties using \autoref{eqn:GaussianFit}.

We assume that the position measurements are Gaussian-distributed, with individual likelihoods of

\begin{equation}
\ell_{x,k} = \frac{1}{\sqrt{2 \pi \left( \sigma_{x,k}^2 + \sigma_x^2 \right)}} \exp\left[ - \frac{1}{2} \frac{\left( x_k - X_k \right)^2}{\sigma_{x,k}^2 + \sigma_x^2} \right]
\end{equation}

\noindent and

\begin{equation}
\ell_{y,k} = \frac{1}{\sqrt{2 \pi \left( \sigma_{y,k}^2 + \sigma_y^2 \right)}} \exp\left[ - \frac{1}{2} \frac{\left( y_k - Y_k \right)^2}{\sigma_{y,k}^2 + \sigma_y^2} \right]
\end{equation}

\noindent for the $x$ and $y$ position measurements, respectively.  Here, we use lowercase letters $(x_k,y_k)$ to denote measured values and capital letters $(X_k,Y_k)$ to denote modeled values.  We have also included ``error floor'' model parameters, $\sigma_x$ and $\sigma_y$, to account for additional sources of measurement uncertainty not captured by \autoref{eqn:GaussianFit}.  Multiplying the probabilities for all individual maser spots yields the joint log-likelihood for the position measurements,

\begin{equation}
\ln\left( \mathcal{L}_1 \right) = - \frac{1}{2} \sum_k \left[ \frac{\left( x_k - X_k \right)^2}{\sigma_{x,k}^2 + \sigma_x^2} + \frac{\left( y_k - Y_k \right)^2}{\sigma_{y,k}^2 + \sigma_y^2} + \ln\left[ 2 \pi \left( \sigma_{x,k}^2 + \sigma_x^2 \right) \right] + \ln\left[ 2 \pi \left( \sigma_{y,k}^2 + \sigma_y^2 \right) \right] \right] . \label{eqn:lnL1}
\end{equation}

We treat the acceleration measurements in a similar manner to the position measurements.  We construct an error floor parameter $\sigma_a$ that gets added in quadrature with the measurement uncertainties $\sigma_{a,k}$.  The probability to measure a maser spot to have acceleration $a_k$ when its ``true" acceleration is $A_k$ is then given by

\begin{equation}
\ell_{a,k} = \frac{1}{\sqrt{2 \pi \left( \sigma_{a,k}^2 + \sigma_a^2 \right)}} \exp\left[ - \frac{1}{2} \frac{\left( a_k - A_k \right)^2}{\sigma_{a,k}^2 + \sigma_a^2} \right] .
\end{equation}

\noindent We multiply the individual probabilities for all maser spots to construct the joint likelihood,

\begin{equation}
\ln\left( \mathcal{L}_2 \right) = - \frac{1}{2} \sum_k \left[ \frac{\left( a_k - A_k \right)^2}{\sigma_{a,k}^2 + \sigma_a^2} + \ln\left[ 2 \pi \left( \sigma_{a,k}^2 + \sigma_a^2 \right) \right] \right] . \label{eqn:lnL2}
\end{equation}

\noindent For those maser features that are too weak to measure accelerations directly, we simply exclude the associated acceleration measurement from \autoref{eqn:lnL2}.

Our velocity ``measurements" are obtained in a qualitatively different manner than either the position or acceleration measurements, and do not come with obvious associated measurement uncertainties.  The velocity $v_k$ we associate with any particular maser spot corresponds to the central velocity of a spectral channel in a VLBI map.  The calibration uncertainties in these velocities are negligible, so the effective uncertainty arises instead because (1) the spectral channels are discretized and finitely wide, (2) a maser line may span more than one channel, and (3) maser lines may have drifted in velocity over the $\sim$4-month interval between the first and last VLBI observation.  While the position and acceleration measurements are made using centroiding techniques and thus are continuous across their measurement domains, the velocity measurements take on discretized values in integer multiples of the channel width, with offsets determined by the specific spectral gridding scheme.  This discretization will necessarily introduce some uncertainty into the velocity measurements, though the exact value and form that this uncertainty should take is not obvious.  An additional source of uncertainty arises for maser lines that span multiple spectral channels, in which case the velocity of any one channel is not necessarily reflective of the true maser velocity. \\ \vspace{-2mm}

We determine the velocity uncertainties by introducing two error floor parameters, $\sigma_{v,\text{sys}}$ and $\sigma_{v,\text{hv}}$, that describe the systemic and high-velocity features, respectively.  Constructing the probability,

\begin{equation}
\ell_{v,k} = 
\begin{dcases}
\frac{1}{\sqrt{2 \pi \sigma_{v,\text{sys}}^2}} \exp\left( - \frac{\left( v_k - V_k \right)^2}{2 \sigma_{v,\text{sys}}^2} \right) & \text{for systemic features} \\
\frac{1}{\sqrt{2 \pi \sigma_{v,\text{hv}}^2}} \exp\left( - \frac{\left( v_k - V_k \right)^2}{2 \sigma_{v,\text{hv}}^2} \right) & \text{for high-velocity features}
\end{dcases} ,
\end{equation}

\noindent and then multiplying these individual probabilities for all maser spots yields the likelihood,

\begin{equation}
\ln\left( \mathcal{L}_3 \right) = 
\begin{dcases}
- \frac{1}{2} \sum_k \left[ \frac{\left( v_k - V_k \right)^2}{\sigma_{v,\text{sys}}^2} + \ln\left( 2 \pi \sigma_{v,\text{sys}}^2 \right) \right] & \text{for systemic features} \\
- \frac{1}{2} \sum_k \left[ \frac{\left( v_k - V_k \right)^2}{\sigma_{v,\text{hv}}^2} + \ln\left( 2 \pi \sigma_{v,\text{hv}}^2 \right) \right] & \text{for high-velocity features}
\end{dcases} . \label{eqn:lnL3}
\end{equation}

Having constructed the likelihood functions for all three measurement classes, we combine them to obtain the overall likelihood function,

\begin{equation}
\ln\left( \mathcal{L} \right) = \ln\left( \mathcal{L}_1 \right) + \ln\left( \mathcal{L}_2 \right) + \ln\left( \mathcal{L}_3 \right) . \label{eqn:LnL}
\end{equation}

\noindent The final model contains 16 global parameters: $D$, $M_{\text{BH}}$, $z_0$, $x_0$, $y_0$, $i_0$, $\frac{di}{dr}$, $\Omega_0$, $\frac{d\Omega}{dr}$, $\omega_0$, $\frac{d\omega}{dr}$, and the error floor parameters $\sigma_x$, $\sigma_y$, $\sigma_a$, $\sigma_{v,\text{sys}}$, and $\sigma_{v,\text{hv}}$.  It also contains many other nuisance parameters associated with the positions of individual maser spots and the accelerations of the weakest spots.  For a fit to $N_r$, $N_b$, and $N_s$ redshifted, blueshifted, and systemic maser spots, respectively, the model will have $2 \left( N_r + N_b + N_s \right)$ additional free parameters corresponding to a $(r,\phi)$ pair for every maser feature.  Similarly, for $N_a$ maser features with fitted (rather than measured) accelerations, the model gains an additional $N_a$ free parameters.  For this work, $N_r = 71$, $N_b = 50$, and $N_s = 45$, bringing the total number of model parameters to 348.  The number of measurements is $4(N_r + N_b + N_s)$ minus the number $N_a = 20$ of unmeasured accelerations, for a total of 604 constraints and 256 degrees of freedom.

We use only uniform and Normal (i.e., Gaussian) priors in this paper.  For any particular parameter $\Theta$, the uniform prior is given by

\begin{equation}
\mathcal{U}(a,b) = \begin{cases}
\frac{1}{b-a} & a \leq \Theta \leq b \\
0 & \text{otherwise}
\end{cases} , \label{eqn:UniformPrior}
\end{equation}

\noindent and the Normal prior is given by

\begin{equation}
\mathcal{N}(\mu,\sigma) = \frac{1}{\sqrt{2 \pi \sigma^2}} \exp\left[ -\frac{1}{2} \left( \frac{\Theta - \mu}{\sigma} \right)^2 \right] . \label{eqn:NormalPrior}
\end{equation}

\noindent For most fits we assign uniform priors to all parameters. \autoref{tab:DiskFittingResults} lists the priors for all non-nuisance parameters.  Both $\sigma_x$ and $\sigma_y$ are assigned uniform priors on the range $[0,0.1]$\,mas, $\sigma_a$ is assigned a uniform prior on $[0,20]$\,\kmsyr, and  $\sigma_{v,\text{sys}}$ and $\sigma_{v,\text{hv}}$ are assigned uniform priors on the range $[0,20]$\,\kms.  All $r$ parameters for the maser spots are assigned uniform priors within the range $[0.1,1.5]$~mas.  The $\phi$ parameters are assigned uniform priors in the range $[0,\pi]$ for the redshifted features, $[\pi,2\pi]$ for the blueshifted features, and $[-\frac{\pi}{2} , \frac{\pi}{2}]$ for the systemic features.

For the majority of the model parameters, our chosen prior distributions are significantly broader than the associated posterior distributions.  The only parameters for which we have found that the prior plays a non-negligible role compared to the data are the error floor parameters, for which we explore the impact of prior assumptions as a component of the systematic error budget (see \autoref{sec:FittingProcedure}).

\bibliography{main}{}
\bibliographystyle{aasjournal}

\end{document}